\documentclass[aps,preprint]{revtex4}%
\usepackage{amsfonts}
\usepackage{amsmath}
\usepackage{amssymb}
\usepackage{bm}
\usepackage{graphicx}%
\begin{document}
\preprint{ }

\vspace*{1cm}

\begin{center}
{\Large Lattice Simulations for Light Nuclei: \\ Chiral Effective Field Theory
at Leading Order}\vspace*{1cm}


{Bu\={g}ra~Borasoy$^{a}$, Evgeny Epelbaum$^{a,b}$, Hermann~Krebs$^{a,b}$,
Dean~Lee$^{c}$, Ulf-G.~Mei{\ss }ner$^{a,b}$}\vspace*{0.75cm}

$^{a}$\textit{Helmholtz-Institut f\"{u}r Strahlen- und Kernphysik (Theorie),
Universit\"{a}t Bonn,\\ Nu\ss allee 14-16, D-53115 Bonn, Germany }\\[0.1cm]

$^{b}$\textit{Institut f\"{u}r Kernphysik (Theorie), Forschungszentrum
J\"{u}lich, D-52425 J\"{u}lich, Germany }

$^{c}$\textit{Department of Physics, North Carolina State University, Raleigh,
NC 27603, USA}

\vspace*{0.75cm}

{\large Abstract}
\end{center}

We discuss lattice simulations of light nuclei at leading order in chiral
effective field theory. \ Using lattice pion fields and auxiliary fields, we
include the physics of instantaneous one-pion exchange and the leading-order
S-wave contact interactions. \ We also consider higher-derivative contact
interactions which adjust the S-wave scattering amplitude at higher momenta.
\ By construction our lattice path integral is positive definite in the limit
of exact Wigner $SU(4)$ symmetry for any even number of nucleons.  
\ This $SU(4)$ positivity and the
approximate $SU(4)$\ symmetry of the low-energy interactions play an important
role in suppressing sign and phase oscillations in Monte Carlo simulations.
\ We assess the computational scaling of the lattice algorithm for light
nuclei with up to eight nucleons and analyze in detail calculations of the
deuteron, triton, and helium-4.

\vspace*{0.75cm}

\section{Introduction}

The underlying theory of strong interactions, quantum chromodynamics (QCD),
describes the interactions of quarks and gluons. \ While analytic calculations
of the properties of confined quarks and gluons inside hadrons are not
possible, a model-independent way of calculating observables directly from QCD
is provided by lattice field theory. \ Recent advances in lattice QCD have
made it possible to calculate the spectrum and properties of various
isolated hadrons. \ There has also been progress in calculating low-energy
hadronic interactions such as pion-pion scattering
\cite{Kuramashi:1993ka,Aoki:2002ny,Lin:2002aj,Beane:2005rj}.\ \ Other hadronic
interactions such as nucleon-nucleon scattering are more difficult, but there
has been some promising recent work in this direction as well
\cite{Fukugita:1994ve,Beane:2003da,Beane:2006mx}.

Unfortunately lattice QCD calculations of many-body systems of nuclear and
neutron matter or even few-body systems beyond two nucleons are presently out
of reach. \ Such simulations would require pion masses at or near the physical
mass and lattices several times longer in each dimension than used in current
simulations. \ But the greatest challenge would be to overcome the
exponentially small signal-to-noise ratio for simulations at large quark
number. \ For many-body systems this is manifested as complex phase
oscillations when adding a quark chemical potential. \ For few-body systems
the calculation can be done at zero chemical potential by measuring
correlation functions involving $3A$-quark operators, where $A$ is the number
of nucleons. \ However here the signal-to-noise problem reappears in the
antisymmetrization over quarks and in the small
overlap between Monte Carlo configurations for the QCD vacuum versus the
$A$-nucleon ground state.

For few- and many-body systems in low-energy nuclear physics one can make
further progress by working directly with hadronic degrees of freedom. \ There
are several possible choices for the form of the nuclear forces and the
computational methods 
used to describe the interactions of low-energy protons
and neutrons.

\ For systems with four or fewer nucleons, a 
numerically exact
approach is provided by the Faddeev-Yakubovsky integral equations. 
Three-nucleon continuum observables as well as the triton and
  $\alpha$-particle binding energies \cite{Nogga:2001cz} were extensively
  studied within the Faddeev-Yakubovsky scheme based on a variety of modern
  semi-phenomenological nucleon-nucleon potential models including the CD-Bonn
   \cite{Machleidt:1996km}, CD--Bonn 2000 \cite{Machleidt:2000ge}, Argonne V18 \cite{Wiringa:1995wb} and Nijmegen
   \cite{Stoks:1994wp} potentials. Three-nucleon forces were also
   incorporated using the Tucson-Melbourne \cite{Coon:1981TM,Coon:2001pv}, Urbana-IX
   \cite{Pudliner:1997ck} and other models. For a
   comprehensive review on the calculations in the three-nucleon continuum the
  reader is referred to \cite{Gloeckle:1995jg}. The same computational scheme
  was applied to nuclear forces derived in chiral effective field theory (ChEFT)
  both at next-to-leading order (NLO) \cite{Epelbaum:2000mx} and at next-to-next-to-leading order
  (NNLO) \cite{Epelbaum:2002vt} in the chiral expansion. 
  Applications of the low-momentum interaction potential
$V_{\text{low }k}$ \cite{Epelbaum:1998hg,Bogner:2001gq,Bogner:2003wn} to few-nucleon systems
  are considered in Refs.~\cite{Fujii:2004dd,Nogga:2004ab}. 
 \ Further computational techniques such as, e.g., the expansion in hyperspherical
  harmonics \cite{Kievsky:1992um}, the Lorentz integral transform method \cite{Efros:1994iq},
  the stochastic variational method \cite{Varga:1995dm} and the
  Kohn-variational approach \cite{Kievsky:1997zf} were also applied to few-nucleon systems. 

For systems with more nucleons one must rely on techniques such as Monte Carlo
simulations or basis-truncated eigenvector methods. \ There have been a number
of Green's Function Monte Carlo simulations of light nuclei based on AV18 as
well as other phenomenological potentials, see for example
\cite{Pudliner:1997ck,Wiringa:2000gb,Pieper:2001mp,Pieper:2001ap,Pieper:2002ne,Wiringa:2002ja,Pieper:2004qw}%
. \ A related technique implementing diffusion Monte Carlo with auxiliary
fields has been used to study the ground state of neutron matter and neutron
droplets \cite{Fantoni:2001ih,Sarsa:2003zu,Pederiva:2004iz,Chang:2004sj}.
\ The No-Core Shell Model (NCSM) is a different approach to light nuclei which
uses basis-truncated eigenvector methods. \ There have been several NCSM
calculations using various different phenomenological potential models, cf.
\cite{Navratil:2000ww,Fayache:2001kq,Navratil:2003ef,Caurier:2005rb}. \ There
are also NCSM calculations which have used nuclear forces derived from ChEFT
\cite{Forssen:2004dk,Nogga:2005hp}. \ Quite recently there has been work in
constructing a low-energy effective field theory within the framework of
truncated basis states used in the NCSM formalism \cite{Stetcu:2006ey}. \ A
benchmark comparison of many of the methods listed above as well as other
techniques can be found in \cite{Kamada:2001tv}.
A review article on various
methods used for light nuclei can be found in \cite{Carlson:qn}.

In this study we consider nuclear lattice simulations of light nuclei using
chiral effective theory. \ The nuclear lattice approach addresses the few- and
many-body problem in nuclear physics by applying non-perturbative lattice
methods to low-energy nucleons and pions. \ The chiral effective Lagrangian is
formulated on a Euclidean lattice and the path integral is evaluated by Monte
Carlo sampling. \ Pions and nucleons are treated as point-like particles on
the lattice sites, and $\pi$ times the inverse lattice spacing sets the cutoff
scale in momentum space. \ By using hadronic degrees of freedom and
concentrating on low-energy physics, it is possible to probe larger volumes,
lower temperatures, and far greater numbers of nucleons than in lattice QCD.
\ In some cases the sign and complex phase oscillations in Monte Carlo
simulations can be significantly reduced or even completely eliminated.

The first study combining lattice methods with effective field theory for
low-energy nuclear physics looked at infinite nuclear and neutron matter at
nonzero density and temperature \cite{Muller:1999cp}. \ The approach we use
here is based on chiral effective field theory starting at leading order.
\ This lattice formalism was also used in \cite{Lee:2004si} to study neutron
matter at nonzero temperature. \ We list some features of the nuclear lattice
approach which seem promising and distinguish it from\ other few- and
many-body techniques.

One unique feature of the lattice effective field theory approach is the
ability to study in the same formalism both few- and many-body systems as well as
zero- and nonzero-temperature phenomena. \ A large portion of the nuclear
phase diagram can be studied using exactly the same lattice action with
exactly the same operator coefficients. \ A second feature is the
computational advantage of many efficient Euclidean lattice methods developed
for lattice QCD and condensed matter applications. \ This includes the use of
Markov Chain Monte Carlo techniques, Hubbard-Stratonovich transformations
\cite{Hubbard:1959ub,Stratonovich:1958}, and non-local updating schemes such
as a hybrid Monte Carlo \cite{Duane:1987de}. \ A third feature is the close
theoretical link between nuclear lattice simulations and chiral effective
field theory. \ 
One can write down the lattice Feynman rules and calculate
lattice Feynman diagrams using precisely the same action used in the
non-perturbative simulation. \ Since the lattice formalism is based on chiral
effective field theory we have a systematic power-counting expansion, an a
priori estimate of errors for low-energy scattering, and a clear theoretical
connection to the underlying symmetries of QCD.

Nuclear lattice simulations were used to study the triton at leading-order in
pionless effective field theory with three-nucleon interactions
\cite{Borasoy:2005yc}. \ In the present investigation we consider the physics
of instantaneous one-pion exchange and the leading-order S-wave contact
interactions. \ We also consider higher-derivative contact interactions which
adjust the S-wave scattering amplitude at higher momenta. \ We calculate
binding energies, radii, and density correlations for the deuteron, triton,
and helium-4, and probe the computational scaling in systems with up to eight nucleons.

\section{Chiral effective field theory in the few-nucleon sector}

Chiral perturbation theory in the purely mesonic sector has a rigorous chiral
counting scheme. \ In the one-nucleon sector a chiral counting scheme can be
established by various means such as the heavy-baryon formulation
\cite{Jenkins:1990jv, Bernard:1992qa} or infrared regularization
\cite{Becher:1999he}. \ In each case Green's functions are expanded in
increasing powers of pion masses and small momenta, and the chiral expansion
corresponds to a loop expansion.

In the few-nucleon sector however one has to deal with non-perturbative
effects. \ Perturbation theory fails at low energies in the few-nucleon sector
due to enhanced contributions from reducible diagrams which contain purely
nucleonic intermediate states. In order to circumvent this problem, one
derives first the interaction kernel (or effective potential) from all
possible irreducible terms without purely nucleonic intermediate states
\cite{Weinberg:1990rz, Weinberg:1991um}. The interaction kernel does not
contain small energy denominators and obeys the conventional chiral counting
scheme. \ The Green's function is then obtained by iterating the interaction
kernel to infinite order in a bound state or scattering state equation.

At lowest order in the chiral expansion the effective Lagrange density is%
\begin{align}
\mathcal{L}  &  =\frac{1}{2}\partial_{\mu}\boldsymbol\pi\cdot\partial^{\mu
}\boldsymbol\pi-\frac{1}{2}m_{\pi}^{2}\boldsymbol\pi^{2}+N^{\dagger}%
i\partial_{0}N+N^{\dagger}\frac{\vec{\nabla}^{2}}{2m}N\nonumber\\
&  -\frac{g_{A}}{2f_{\pi}}N^{\dagger}\boldsymbol\tau\ \vec{\sigma}\cdot
\cdot\vec{\nabla}\boldsymbol\pi N-\frac{1}{2}C(N^{\dagger}N)(N^{\dagger
}N)-\frac{1}{2}C_{I}(N^{\dagger}\boldsymbol\tau N)\cdot(N^{\dagger
}\boldsymbol\tau N). \label{effectiveL}%
\end{align}
\ We use the same notation as used in \cite{Epelbaum:thesis}. \ $N$ is the
nucleon field with spin and isospin degrees of freedom. \ The vector arrow in
$\vec{\sigma}$ signifies the three-vector index for spin. \ The boldface for
$\boldsymbol\tau$ and $\boldsymbol\pi$ signifies the three-vector index for
isospin. \ We take for our physical constants $m=938.92$ MeV as the nucleon
mass, $m_{\pi}=138.08$ MeV as the pion mass, $f_{\pi}=93$~MeV as the pion
decay constant, and $g_{A}=1.26$ as the nucleon axial charge. \ In order to
reduce sign and complex phase oscillations in the Monte Carlo calculation with
auxiliary fields (see \cite{Lee:2004ze}) we work with the leading-order
contact interactions $C$ and $C_{I}$ rather than the more standard interaction
coefficients $C_{S}$ and $C_{T}$ corresponding with%
\begin{equation}
-\frac{1}{2}C_{S}(N^{\dagger}N)(N^{\dagger}N)-\frac{1}{2}C_{T}(N^{\dagger}%
\vec{\sigma}N)\cdot(N^{\dagger}\vec{\sigma}N).
\end{equation}
Both forms for the interactions are exactly the same if we set%
\begin{equation}
C=C_{S}-2C_{T},\qquad C_{I}=-C_{T}.
\end{equation}

From the effective Lagrangian in (\ref{effectiveL}) the $NN$ effective
potential can be derived by applying the method of unitary transformations
\cite{Epelbaum:1998ka}, Q-box expansion \cite{Krebs:2004st}, or
other techniques \cite{Ordonez:1996rz, Kaiser:1997mw}. At leading order (LO) 
the $NN$ effective potential consists of
the two contact interactions and instantaneous one-pion exchange,%
\begin{equation}
V_{LO}=C+C_{I}\ \boldsymbol\tau_{1}\cdot\boldsymbol\tau_{2}-\left(  \frac{g_{A}}{2f_{\pi}}\right)
^{2}\boldsymbol\tau_{1}\cdot\boldsymbol\tau_{2}\frac{\vec{\sigma_{1}}\cdot
\vec{q}\ \ \vec{\sigma_{2}}\cdot\vec{q}}{\vec{q}\,^{2}+m_{\pi}^{2}},
\end{equation}
where $\vec q = \vec p \, ' - \vec p$ is the nucleon momentum transfer.  
We can reproduce the desired iteration of $V_{LO}$ if we start with the
Lagrange density,%
\begin{align}  
\mathcal{L}  &  =-\frac{1}{2}\vec{\nabla}\boldsymbol\pi\cdot\cdot\vec{\nabla
}\boldsymbol\pi-\frac{1}{2}m_{\pi}^{2}\boldsymbol\pi^{2}+N^{\dagger}%
i\partial_{0}N+N^{\dagger}\frac{\vec{\nabla}^{2}}{2m}%
N\nonumber\label{eq:modlagr}\\[0.3cm]
&  -\frac{g_{A}}{2f_{\pi}}N^{\dagger}\boldsymbol\tau\ \vec{\sigma}\cdot
\cdot\vec{\nabla}\boldsymbol\pi N-\frac{1}{2}C(N^{\dagger}N)(N^{\dagger
}N)-\frac{1}{2}C_{I}(N^{\dagger}\boldsymbol\tau N)\cdot(N^{\dagger
}\boldsymbol\tau N),
\end{align}
and evaluate 
the $NN$ scattering amplitude
non-perturbatively. \ We note that the pions have no time
derivatives. \ Therefore they can only be exchanged instantaneously between
nucleons and do not propagate in time. Clearly the Lagrangian in Eq.~(\ref{eq:modlagr}) is not valid
for external pion fields. 
The two-nucleon Green's function derived from the path integral representation  
with this Lagrangian reproduces the solution of the corresponding Lippmann-Schwinger
equation with the leading order effective potential.
\ Another advantage of treating pions this
way is that the nucleon self-energy exactly vanishes and the nucleon mass is
not renormalized.

\section{Lattice notation}

In this study we assume exact isospin symmetry and neglect electromagnetic
interactions. \ We use $\vec{n}$ to represent integer-valued coordinate
lattice vectors on a $3+1$ dimensional space-time lattice and $\vec{k}$ to
represent integer-valued momentum lattice vectors.$\ \ $A subscripted `$s$'
such as in $\vec{n}_{s}$ represents purely spatial lattice vectors. \ $\hat
{0}$ denotes the unit lattice vector in the time direction, and $\hat{l}%
_{s}=\hat{1}$, $\hat{2}$, $\hat{3}$ are unit lattice vectors in the spatial
directions. $\ a$ is the spatial lattice spacing, $L$ is the length of the
cubic spatial lattice in each direction, $a_{t}$ is the lattice spacing in the
temporal direction, and $L_{t}$ is the length in the temporal direction. \ We
define $\alpha_{t}$ as the ratio between lattice spacings, $\alpha_{t}%
=a_{t}/a$, and define $h=\alpha_{t}/(2m)$. \ Throughout we use dimensionless
parameters and operators, which correspond with physical values multiplied by
the appropriate power of $a$. \ Final results are presented in physical units
with the corresponding unit stated explicitly.

To avoid confusion we make explicit in our lattice notation all spin and
isospin indices. \ We use $c$ and $c^{\ast}$ to denote the anticommuting
Grassmann variables for the nucleons and $a$ and $a^{\dagger}$ to denote
annihilation and creation operators. \ We use the subscript notation%
\begin{equation}
\left[
\begin{array}
[c]{c}%
c_{\uparrow,p}\\
c_{\downarrow,p}\\
c_{\uparrow,n}\\
c_{\downarrow,n}%
\end{array}
\right]  =\left[
\begin{array}
[c]{c}%
c_{0,0}\\
c_{1,0}\\
c_{0,1}\\
c_{1,1}%
\end{array}
\right]  ,\qquad\left[
\begin{array}
[c]{c}%
a_{\uparrow,p}\\
a_{\downarrow,p}\\
a_{\uparrow,n}\\
a_{\downarrow,n}%
\end{array}
\right]  =\left[
\begin{array}
[c]{c}%
a_{0,0}\\
a_{1,0}\\
a_{0,1}\\
a_{1,1}%
\end{array}
\right]  .
\end{equation}
The first subscript is for spin and the second subscript is for isospin. \ We
use $\tau_{I}$ with $I=1,2,3$ to represent Pauli matrices acting in isospin
space and $\sigma_{S}$ with $S=1,2,3$ to represent Pauli matrices acting in
spin space. \ We note that on the lattice the spin symmetry is reduced to the
cubic subgroup $SO(3,\mathbb{Z})$ of $SO(3)$ while isospin symmetry remains
intact as the full $SU(2)$ symmetry.

\section{Path integral for free nucleons and instantaneous pions}  

\begin{figure}
[bpt]
\begin{center}
\includegraphics[
height=3.6288in,
width=3.3857in
]%
{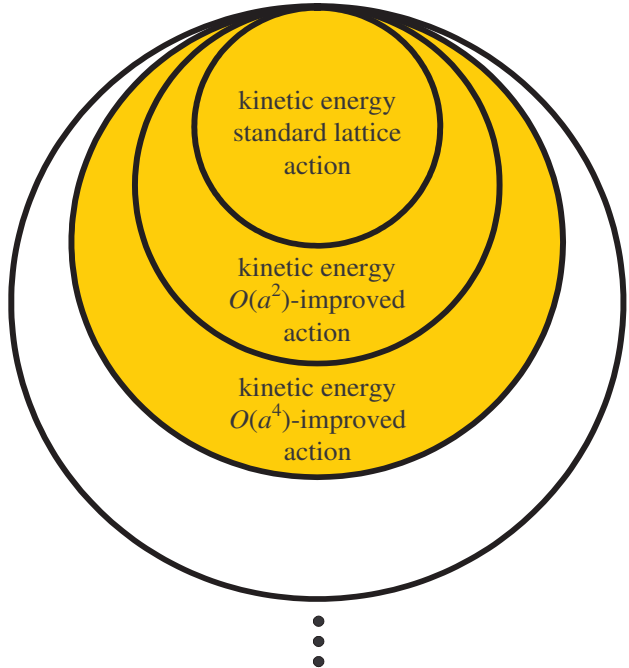}%
\caption{Suppression of finite cutoff errors by introducing improved actions for the nucleon kinetic energy.}%
\label{fig:kineticenergy}%
\end{center}
\end{figure}

If we could take the continuum limit, the accuracy of our calculation would be determined by 
the order $k$ we choose to truncate the chiral expansion.  This would correspond with 
$(p/\Lambda_\chi)^k$, where $p$ is a typical 
low-energy momentum scale and $\Lambda_\chi = 4 \pi f_\pi \simeq 1.2$~GeV the scale of spontaneous 
chiral symmetry breakdown.
However taking the continuum limit is not possible due to the non-perturbative treatment of the 
chiral effective Lagrangian on the lattice as this would require an infinite set
of counterterms. The lattice cutoff $\Lambda$ must be chosen to remain below the scale $\Lambda_\chi$.
This in turn introduces an error of the general form $(p/\Lambda)^{k_1}(\Lambda/\Lambda_\chi)^{k_2}$
due to the finite cutoff and missing counterterms. 
Even for the free nucleon case finite cutoff errors occur which can be traced back
to the discretized lattice propagator.  In this analysis we 
use an ${\cal O}(a^4)$-improved action for 
the lattice kinetic energy, as shown in Figure~\ref{fig:kineticenergy}.

\begin{figure}
[bpt]
\begin{center}
\includegraphics[
height=3.6288in,
width=3.3857in
]%
{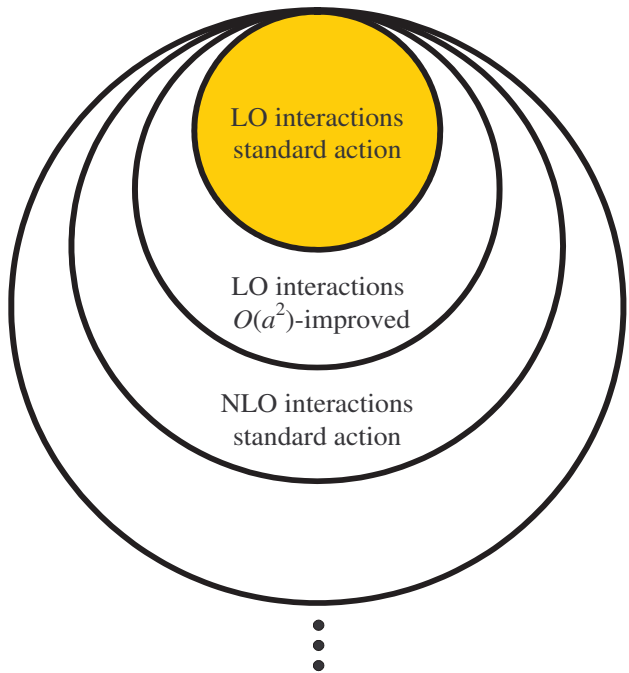}%
\caption{Hierarchy of interactions according to chiral order as well as improvements to each order due 
to nonzero lattice spacing.}%
\label{fig:LOinteractions}%
\end{center}
\end{figure}

Similarly, the interactions in the continuum limit can be organized in the chiral expansion as
leading order, next-to-leading order, 
etc.  But again for a chosen chiral order we may wish to include additional improvements to the 
interactions which reduce the finite cutoff errors.  In this analysis we start 
by considering the simple LO
lattice action without improvement, as shown in Figure~\ref{fig:LOinteractions}.  We note that 
the diagram shown is a bit simplistic since the improvement terms may in general include corrections for 
effects at nonzero lattice spacing such as broken Galilean invariance, etc.

Throughout our discussion we consider both the path integral formalism and the
transfer matrix formalism. \ The path integral formalism is useful for
deriving the lattice Feynman rules and auxiliary field formulations, while the
transfer matrix is used for the Monte Carlo simulations of light nuclei. \ We
start with the path integral formalism. \ Let $\mathcal{Z}_{\bar{N}N}$ be the
lattice partition function for free nucleons%
\begin{equation}
\mathcal{Z}_{\bar{N}N}\propto\int DcDc^{\ast}\exp\left[  -S_{\bar{N}N}\left(
c,c^{\ast}\right)  \right]  ,
\end{equation}
where%
\begin{equation}
DcDc^{\ast}=\prod_{\vec{n},i,j}dc_{i,j}(\vec{n})dc_{i,j}^{\ast}(\vec{n}).
\end{equation}
We use an $O(a^{4})$-improved lattice action with next-to-next-to-nearest
neighbor hopping,%
\begin{align}
S_{\bar{N}N}(c,c^{\ast})=  &  \sum_{\vec{n},i,j}\left[  c_{i,j}^{\ast}(\vec
{n})c_{i,j}(\vec{n}+\hat{0})+\left(  -1+\frac{49h}{6}\right)  c_{i,j}^{\ast
}(\vec{n})c_{i,j}(\vec{n})\right] \nonumber\\
&  -\frac{3h}{2}\sum_{\vec{n},l_{s},i,j}\left[  c_{i,j}^{\ast}(\vec{n}%
)c_{i,j}(\vec{n}+\hat{l}_{s})+c_{i,j}^{\ast}(\vec{n})c_{i,j}(\vec{n}-\hat
{l}_{s})\right] \nonumber\\
&  +\frac{3h}{20}\sum_{\vec{n},l_{s},i,j}\left[  c_{i,j}^{\ast}(\vec
{n})c_{i,j}(\vec{n}+2\hat{l}_{s})+c_{i,j}^{\ast}(\vec{n})c_{i,j}(\vec{n}%
-2\hat{l}_{s})\right] \nonumber\\
&  -\frac{h}{90}\sum_{\vec{n},l_{s},i,j}\left[  c_{i,j}^{\ast}(\vec{n}%
)c_{i,j}(\vec{n}+3\hat{l}_{s})+c_{i,j}^{\ast}(\vec{n})c_{i,j}(\vec{n}-3\hat
{l}_{s})\right]  .
\end{align}
We expect the $O(a^{4})$-improvement in the lattice dispersion relation 
to be useful when measuring scattering phase shifts on the lattice.

In this leading-order study we consider instantaneous one-pion exchange and no
other interactions involving pions. \ In our lattice formalism the pion field
does not propagate in time and does not couple to physical pions. \ This
allows to avoid
the problem of non-perturbative dynamical pion fields producing
unrenormalized pion loops to all orders. \ If at some point later on we wish
to include interactions with physical low-energy pions we simply insert the
corresponding operators with external pion fields.

The lattice action for free pions with purely instantaneous propagation is%
\begin{equation}
S_{\pi\pi}(\pi_{I})=\alpha_{t}(\tfrac{m_{\pi}^{2}}{2}+3)\sum_{I=1,2,3}%
\sum_{\vec{n}}\pi_{I}(\vec{n})\pi_{I}(\vec{n})-\alpha_{t}\sum_{I=1,2,3}%
\sum_{\vec{n},l_{s}}\pi_{I}(\vec{n})\pi_{I}(\vec{n}+\hat{l}_{s}).
\end{equation}
In order to simplify the Monte Carlo updating scheme later in our discussion
it is helpful at this point to define a rescaled pion field,%
\begin{equation}
\pi_{I}^{\prime}(\vec{n})=\sqrt{q_{\pi}}\pi_{I}(\vec{n}),
\end{equation}
where%
\begin{equation}
q_{\pi}=\alpha_{t}(m_{\pi}^{2}+6).
\end{equation}
Then%
\begin{equation}
S_{\pi\pi}(\pi_{I}^{\prime})=\frac{1}{2}\sum_{I=1,2,3}\sum_{\vec{n}}\pi
_{I}^{\prime}(\vec{n})\pi_{I}^{\prime}(\vec{n})-\frac{\alpha_{t}}{q_{\pi}}%
\sum_{I=1,2,3}\sum_{\vec{n},l_{s}}\pi_{I}^{\prime}(\vec{n})\pi_{I}^{\prime
}(\vec{n}+\hat{l}_{s}).
\end{equation}
In momentum space the action is 
\begin{equation}
S_{\pi\pi}(\pi_{I}^{\prime})= \frac{1}{L_{t}L^{3}} \sum_{I=1,2,3}\sum_{\vec{k}}\pi_{I}^{\prime
}(-\vec{k})\pi_{I}^{\prime}(\vec{k})\left[  \frac{1}{2}-\frac{\alpha_{t}%
}{q_{\pi}}\sum_{l_{s}}\cos\left(  \tfrac{2\pi k_{l_{s}}}{L}\right)  \right]  ,
\end{equation}
and so
\begin{equation}
\frac{\int D\pi_{I}^{\prime}\pi_{I}^{\prime}(\vec{n})\pi_{I}^{\prime}(\vec
{0})\exp\left[  -S_{\pi\pi}\right]  }{\int D\pi_{I}^{\prime}\exp\left[
-S_{\pi\pi}\right]  }\text{ (no sum on }I\text{)}=\frac{1}{L_{t}L^{3}}%
\sum_{\vec{k}}e^{-i\frac{2\pi}{L_{t}}k_{t}\cdot n_{t}}e^{-i\frac{2\pi}{L}%
\vec{k}_{s}\cdot\vec{n}_{s}}D_{\pi}(\vec{k}_{s}),
\end{equation}
where the pion propagator is%
\begin{equation}
D_{\pi}(\vec{k}_{s})=\frac{1}{1-\tfrac{2\alpha_{t}}{q_{\pi}}\sum\limits_{l_{s}%
=1,2,3}\cos\left(  \tfrac{2\pi k_{l_{s}}}{L}\right)  }.
\end{equation}
The pion correlation function at spatial separation $\vec{n}_{s}$ is then%
\begin{align}
\left\langle \pi_{I}^{\prime}(\vec{n}_{s})\pi_{I}^{\prime}(\vec{0}%
)\right\rangle  &  =\frac{\int D\pi_{I}^{\prime}\pi_{I}^{\prime}(\vec{n}%
_{s})\pi_{I}^{\prime}(\vec{0})\exp\left[  -S_{\pi\pi}\right]  }{\int D\pi
_{I}^{\prime}\exp\left[  -S_{\pi\pi}\right]  }\text{ (no sum on }%
I\text{)}\nonumber\\
&  =\frac{1}{L^{3}}\sum_{\vec{k}_{s}}e^{-i\frac{2\pi}{L}\vec{k}_{s}\cdot
\vec{n}_{s}}D_{\pi}(\vec{k}_{s}).
\end{align}

\section{Pion-nucleon coupling}

There are various ways to introduce spatial derivatives of the pion field on
the lattice. \ The simplest definition for the gradient of $\pi_{I}^{\prime}$
is to define a forward-backward lattice derivative. \ For example we can write%
\begin{equation}
\partial_{1}\pi_{I}^{\prime}(\vec{n})=\frac{1}{2}\left[  \pi_{I}^{\prime}%
(\vec{n}+\hat{1})-\pi_{I}^{\prime}(\vec{n}-\hat{1})\right]  .
\end{equation}
This is the method used in \cite{Lee:2004si}. \ The disadvantage is that it is
a coarse derivative involving a separation distance of two lattice units. \ We
can avoid this if we think of the pion lattice points as being shifted by
$-1/2$ lattice unit from the nucleon lattice points in each of the three
spatial directions. \ For each nucleon lattice point $\vec{n}_{\text{nucleon}%
}$ we associate a pion lattice point $\vec{n}_{\text{pion}}$,%
\begin{equation}
\vec{n}_{\text{pion}}=\vec{n}_{\text{nucleon}}-\frac{1}{2}\hat{1}-\frac{1}%
{2}\hat{2}-\frac{1}{2}\hat{3}.
\end{equation}
Then we have eight pion lattice points forming a cube centered at $\vec
{n}_{\text{nucleon}}$,%
\begin{align}
&  \vec{n}_{\text{pion}},\quad\vec{n}_{\text{pion}}+\hat{1},\quad\vec
{n}_{\text{pion}}+\hat{2},\quad\vec{n}_{\text{pion}}+\hat{3},\nonumber\\
&  \vec{n}_{\text{pion}}+\hat{1}+\hat{2},\quad\vec{n}_{\text{pion}}+\hat
{2}+\hat{3},\quad\vec{n}_{\text{pion}}+\hat{3}+\hat{1},\quad\vec
{n}_{\text{pion}}+\hat{1}+\hat{2}+\hat{3}. \label{pionlattice}%
\end{align}
We use the same lattice vector notation $\vec{n}$ for both nucleons and pions.
\ However for nucleon fields and auxiliary fields to be introduced later
$\vec{n}$ represents $\vec{n}_{\text{nucleon}}$ while for pion fields $\vec
{n}$ refers to $\vec{n}_{\text{pion}}.$

The eight vertices of the pion cube in (\ref{pionlattice}) can be used to
define spatial derivatives of the pion field. \ For each spatial direction
$S=1,2,3$ we have%
\begin{equation}
\Delta_{S}\pi_{I}^{\prime}(\vec{n})=\frac{1}{4}\sum_{\substack{\nu_{1},\nu
_{2},\nu_{3}=0,1}}(-1)^{\nu_{S}+1}\pi_{I}^{\prime}(\vec{n}+\vec{\nu}%
),\qquad\vec{\nu}=\nu_{1}\hat{1}+\nu_{2}\hat{2}+\nu_{3}\hat{3}.
\end{equation}
The lattice pion-nucleon coupling in our lattice action is%
\begin{equation}
S_{\pi\bar{N}N}(\pi_{I}^{\prime},c,c^{\ast})=\tfrac{g_{A}\alpha_{t}}{2f_{\pi
}\sqrt{q_{\pi}}}%
{\displaystyle\sum_{S,I=1,2,3}}
\Delta_{S}\pi_{I}^{\prime}(\vec{n})\rho_{S,I}(\vec{n}),
\end{equation}
where $\rho_{S,I}(\vec{n})$ is the spin-isospin density,%
\begin{equation}
\rho_{S,I}(\vec{n})=\sum_{i,j,i^{\prime},j^{\prime}=0,1}c_{i,j}^{\ast}(\vec
{n})\left[  \sigma_{S}\right]  _{ii^{\prime}}\left[  \tau_{I}\right]
_{jj^{\prime}}c_{i^{\prime},j^{\prime}}(\vec{n}).
\end{equation}

\section{S-wave contact interactions}

There are two S-wave contact interactions at lowest order. \ Following
\cite{Lee:2004ze} we choose the form%
\begin{equation}
S_{\bar{N}N\bar{N}N}(c,c^{\ast})=\frac{C\alpha_{t}}{2}\sum_{\vec{n}}\left[
\rho(\vec{n})\right]  ^{2}+\frac{C_{I}\alpha_{t}}{2}\sum_{I=1,2,3}\sum
_{\vec{n}}\left[  \rho_{I}(\vec{n})\right]  ^{2}, \label{NNNN_action}%
\end{equation}
where $\rho(\vec{n})$ and $\rho_{I}(\vec{n})$ are the $SU(4)$-symmetric 
and isospin densities respectively,%
\begin{equation}
\rho(\vec{n})=\sum_{i,j=0,1}c_{i,j}^{\ast}(\vec{n})c_{i,j}(\vec{n}),
\end{equation}%
\begin{equation}
\rho_{I}(\vec{n})=\sum_{i,j,j^{\prime}=0,1}c_{i,j}^{\ast}(\vec{n})\left[
\tau_{I}\right]  _{jj^{\prime}}c_{i,j^{\prime}}(\vec{n}).
\end{equation}
Since the isospin singlet channel is more strongly attractive than the isospin
triplet channel, we anticipate the signs for these
coefficients to be $C<0$ and $C_{I}>0$.
This will be confirmed in Sec.~\ref{sec:2N}, where the two-nucleon system is studied in detail.
\ As noted above, these can be written in terms of the more familiar
coefficients $C_{S}$ and $C_{T}$ using the identity%
\begin{equation}
C=C_{S}-2C_{T},\qquad C_{I}=-C_{T}.
\end{equation}

We can use the Gaussian integral identities%
\begin{equation}
\exp\left\{  -\frac{C\alpha_{t}}{2}\left[  \rho(\vec{n})\right]  ^{2}\right\}
=\frac{1}{\sqrt{2\pi}}\int_{-\infty}^{\infty}ds\exp\left[  -\frac{1}{2}%
s^{2}+\sqrt{-C\alpha_{t}}\rho(\vec{n})\cdot s\right]  ,
\end{equation}
and%
\begin{align}
&  \exp\left\{  -\frac{C_{I}\alpha_{t}}{2}\sum_{I=1,2,3}\left[  \rho_{I}%
(\vec{n})\right]  ^{2}\right\} \nonumber\\
&  =\int\left(  \prod\limits_{I=1,2,3}\frac{ds_{I}}{\sqrt{2\pi}}\right)
\exp\left[  -\frac{1}{2}\sum_{I=1,2,3}\left(  s_{I}\right)  ^{2}+i\sqrt
{C_{I}\alpha_{t}}\sum_{I=1,2,3}\rho_{I}(\vec{n})\cdot s_{I}\right]  .
\end{align}
Let us define the auxiliary field actions,%
\begin{equation}
S_{ss}(s,s_{I})=\frac{1}{2}\sum_{\vec{n}}s^{2}(\vec{n})+\frac{1}{2}%
\sum_{I=1,2,3}\sum_{\vec{n}}\left[  s_{I}(\vec{n})\right]  ^{2},
\label{ss_action}%
\end{equation}%
\begin{equation}
S_{s\bar{N}N}(s,s_{I},c,c^{\ast})=-\sqrt{-C\alpha_{t}}\sum_{\vec{n}}s(\vec
{n})\rho(\vec{n})-i\sqrt{C_{I}\alpha_{t}}\sum_{I=1,2,3}\sum_{\vec{n}}%
s_{I}(\vec{n})\rho_{I}(\vec{n}). \label{sNN_action}%
\end{equation}
Then we have%
\begin{equation}
\int DsDs_{I}\;\exp\left[  -S_{ss}(s,s_{I})-S_{s\bar{N}N}(s,s_{I},c,c^{\ast
})\right]  \propto \exp\left[  -S_{\bar{N}N\bar{N}N}(c,c^{\ast})\right]  ,
\end{equation}
where%
\begin{equation}
DsDs_{I}=\prod_{\vec{n},I}ds(\vec{n})ds_{I}(\vec{n}).
\end{equation}
If we put all the pieces together the full path integral action at leading
order is%
\begin{equation}
\mathcal{Z}_{LO}\propto%
{\displaystyle\int}
DcDc^{\ast}D\pi_{I}^{\prime}DsDs_{I}\exp\left[  -S_{LO}(c,c^{\ast},\pi
_{I}^{\prime},s,s_{I})\right]  ,
\end{equation}
where%
\begin{equation}
S_{LO}=S_{\bar{N}N}+S_{\pi\pi}+S_{\pi\bar{N}N}+S_{ss}+S_{s\bar{N}N}.
\label{LO_action}%
\end{equation}

\section{Transfer matrix with auxiliary fields}

The transfer matrix is the analog at nonzero temporal lattice spacing of the
operator $\exp\left(  -H\Delta t\right)  $. \ In order to derive the transfer
matrix corresponding with the path integral action $S_{LO}$ we use the
correspondence \cite{Creutz:1988wv,Creutz:1999zy}%
\begin{align}
&  {\rm Tr}\left\{  \colon F_{L_{t}-1}\left[  a_{i^{\prime},j^{\prime}}^{\dagger
}(\vec{n}_{s}^{\prime}),a_{i,j}(\vec{n}_{s})\right]  \colon\times\cdots
\times\colon F_{0}\left[  a_{i^{\prime},j^{\prime}}^{\dagger}(\vec{n}%
_{s}^{\prime}),a_{i,j}(\vec{n}_{s})\right]  \colon\right\} \nonumber\\
&  =\int DcDc^{\ast}\exp\left\{  \sum_{n_{t}=0}^{L_{t}-1}\sum_{\vec{n}%
_{s},i,j}c_{i,j}^{\ast}(\vec{n}_{s},n_{t})\left[  c_{i,j}(\vec{n}_{s}%
,n_{t})-c_{i,j}(\vec{n}_{s},n_{t}+1)\right]  \right\} \nonumber\\
&  \qquad\qquad\qquad\times\prod_{n_{t}=0}^{L_{t}-1}F_{n_{t}}\left[
c_{i^{\prime},j^{\prime}}^{\ast}(\vec{n}_{s}^{\prime},n_{t}),c_{i,j}(\vec
{n}_{s},n_{t})\right]  , \label{correspondence}%
\end{align}
for general functions $F_i$ and antiperiodic boundary conditions in the time direction,
$c_{i,j}(\vec{n}_{s},L_{t})=-c_{i,j}(\vec{n}_{s},0).$ \ The $\colon  \colon$
symbols in (\ref{correspondence}) denote normal ordering. \ Let us define the
$SU(4)$-symmetric, isospin, and spin-isospin densities written in terms of
creation and annihilation operators,%
\begin{equation}
\rho^{a^{\dagger},a}(\vec{n}_{s})=\sum_{i,j=0,1}a_{i,j}^{\dagger}(\vec{n}%
_{s})a_{i,j}(\vec{n}_{s}),
\end{equation}%
\begin{equation}
\rho_{I}^{a^{\dagger},a}(\vec{n}_{s})=\sum_{i,j,j^{\prime}=0,1}a_{i,j}%
^{\dagger}(\vec{n}_{s})\left[  \tau_{I}\right]  _{jj^{\prime}}a_{i,j^{\prime}%
}(\vec{n}_{s}),
\end{equation}%
\begin{equation}
\rho_{S,I}^{a^{\dagger},a}(\vec{n}_{s})=\sum_{i,j,i^{\prime},j^{\prime}%
=0,1}a_{i,j}^{\dagger}(\vec{n}_{s})\left[  \sigma_{S}\right]  _{ii^{\prime}%
}\left[  \tau_{I}\right]  _{jj^{\prime}}a_{i^{\prime},j^{\prime}}(\vec{n}%
_{s}).
\end{equation}
We also define the $O(a^{4})$-improved free nucleon lattice Hamiltonian%
\begin{align}
H_{\text{free}}  &  =\frac{49}{12m}\sum_{\vec{n}_{s},i,j}a_{i,j}^{\dagger
}(\vec{n}_{s})a_{i,j}(\vec{n}_{s})\nonumber\\
&  -\frac{3}{4m}\sum_{\vec{n}_{s},l_{s},i,j}\left[  a_{i,j}^{\dagger}(\vec
{n}_{s})a_{i,j}(\vec{n}_{s}+\hat{l}_{s})+a_{i,j}^{\dagger}(\vec{n}_{s}%
)a_{i,j}(\vec{n}_{s}-\hat{l}_{s})\right] \nonumber\\
&  +\frac{1}{40m}\sum_{\vec{n}_{s},l_{s},i,j}\left[  a_{i,j}^{\dagger}(\vec
{n}_{s})a_{i,j}(\vec{n}_{s}+2\hat{l}_{s})+a_{i,j}^{\dagger}(\vec{n}%
_{s})a_{i,j}(\vec{n}_{s}-2\hat{l}_{s})\right] \nonumber\\
&  -\frac{1}{180m}\sum_{\vec{n}_{s},l_{s},i,j}\left[  a_{i,j}^{\dagger}%
(\vec{n}_{s})a_{i,j}(\vec{n}_{s}+3\hat{l}_{s})+a_{i,j}^{\dagger}(\vec{n}%
_{s})a_{i,j}(\vec{n}_{s}-3\hat{l}_{s})\right]  .
\end{align}
\newline

Using (\ref{correspondence}) the path integral with auxiliary fields can be
expressed in the transfer matrix formalism as%
\begin{equation}
\mathcal{Z}_{LO}\propto%
{\displaystyle\int}
D\pi_{I}^{\prime}DsDs_{I}\;\exp\left[  -S_{\pi\pi}-S_{ss}\right]  \times
Tr\left\{  M^{(L_{t}-1)}(\pi_{I}^{\prime},s,s_{I})\times\cdots\times
M^{(0)}(\pi_{I}^{\prime},s,s_{I})\right\}  ,
\end{equation}
where%
%
\begin{eqnarray}
&& M^{(n_{t})}(\pi_{I}^{\prime},s,s_{I}) 
= \ \colon\exp\left\{
-H_{\text{free}}\alpha_{t}-\tfrac{g_{A}\alpha_{t}}{2f_{\pi}\sqrt{q_{\pi}}}%
{\displaystyle\sum_{S,I}}
\Delta_{S}\pi_{I}^{\prime}(\vec{n}_{s},n_{t})\rho_{S,I}^{a^{\dag},a}(\vec
{n}_{s})\right. \nonumber\\
&& \left.\qquad +\sqrt{-C\alpha_{t}} \sum_{\vec{n}_{s}}s(\vec{n}_{s},n_{t})\rho^{a^{\dag}%
,a}(\vec{n}_{s})+i\sqrt{C_{I}\alpha_{t}}\sum_{I}\sum_{\vec{n}_{s}}s_{I}%
(\vec{n}_{s},n_{t})\rho_{I}^{a^{\dag},a}(\vec{n}_{s})
\right\}  \colon. \label{transfer_matrix}%
\end{eqnarray}

\section{The two-nucleon system} \label{sec:2N}

For the two-nucleon system the entire linear space is small enough for typical
lattice volumes that we can find the low-energy eigenstates on the lattice
using iterative sparse matrix eigenvector methods such as the Lanczos method
\cite{Lanczos:1950}. \ To do this calculation we construct the transfer matrix
with only nucleon fields. \ It is convenient to define
\begin{align}
G_{S_{1}S_{2}}(\vec{n}_{s})  &  =\frac{\int D\pi_{I}^{\prime}\Delta_{S_{1}}%
\pi_{I}^{\prime}(\vec{n}_{s})\Delta_{S_{2}}\pi_{I}^{\prime}(0)\exp\left[
-S_{\pi\pi}\right]  }{\int D\pi_{I}^{\prime}\exp\left[  -S_{\pi\pi}\right]
}\text{ (no sum on }I\text{)}\nonumber\\
&  =\frac{1}{16}\sum_{\nu_{1},\nu_{2},\nu_{3}=0,1}\sum_{\nu_{1}^{\prime}%
,\nu_{2}^{\prime},\nu_{3}^{\prime}=0,1}(-1)^{\nu_{S_{1}}}(-1)^{\nu_{S_{2}%
}^{\prime}}\left\langle \pi_{I}^{\prime}(\vec{n}_{s}+\vec{\nu}-\vec{\nu
}^{\prime})\pi_{I}^{\prime}(\vec{0})\right\rangle .
\end{align}
The path integral can now be written as
\begin{equation}
\mathcal{Z}_{LO}\propto {\rm Tr}\left\{  M^{(L_{t}-1)}\times\cdots\times
M^{(0)}\right\}  ,
\end{equation}
where%
%
\begin{eqnarray}
M^{(n_{t})} &=& \ \colon\exp\left\{
-H_{\text{free}}\alpha_{t}-\frac{1}{2}C\alpha_{t}\sum_{\vec{n}_{s}}\left[
\rho^{a^{\dag},a}(\vec{n}_{s})\right]  ^{2}-\frac{1}{2}C_{I}\alpha_{t}\sum
_{I}\sum_{\vec{n}_{s}}\left[  \rho_{I}^{a^{\dag},a}(\vec{n}_{s})\right]
^{2} \right. \nonumber \\
&& \qquad \qquad \left. +\tfrac{g_{A}^{2}\alpha_{t}^{2}}{8f_{\pi}^{2}q_{\pi}}\sum_{\substack{S_{1}%
,S_{2},I}}\sum_{\vec{n}_{s,1},\vec{n}_{s,2}}G_{S_{1}S_{2}}(\vec{n}_{s,1}%
-\vec{n}_{s,2})\rho_{S_{1},I}^{a^{\dag},a}(\vec{n}_{s,1})\rho_{S_{2}%
,I}^{a^{\dag},a}(\vec{n}_{s,2})
\right\}  \colon. \label{transfer_nucleon_only}%
\end{eqnarray}

We now calculate the two-nucleon spectrum in a periodic cube of length $L$ and
use this information to determine the contact interaction coefficients $C$ and
$C_{I}$. \ We make use of L\"{u}scher's formula
\cite{Luscher:1986pf,Beane:2003da,Seki:2005ns} which relates the two-particle
energy levels in a periodic cube of length $L$ to the S-wave phase shift,%
\begin{equation}
p\cot\delta_{0}(p)=\frac{1}{\pi L}S\left(  \eta\right)  ,\qquad\eta=\left(
\frac{Lp}{2\pi}\right)  ^{2},
\end{equation}
where $S(\eta)$ is the three-dimensional zeta function,%
\begin{equation}
S(\eta)=\lim_{\Lambda\rightarrow\infty}\left[  \sum_{\vec{n}}\frac
{\theta(\Lambda^{2}-\vec{n}^{2})}{\vec{n}^{2}-\eta}-4\pi\Lambda\right]  .
\end{equation}
For $\left\vert \eta\right\vert <1$ we can expand in powers of $\eta$,%
\begin{align}
S(\eta)  &  =-\frac{1}{\eta}+\lim_{\Lambda\rightarrow\infty}\left[  \sum
_{\vec{n}\neq\vec{0}}\frac{\theta(\Lambda^{2}-\vec{n}^{2})}{\vec{n}^{2}-\eta
}-4\pi\Lambda\right] \\
&  =-\frac{1}{\eta}+S_{0}+S_{1}\eta^{1}+S_{2}\eta^{2}+S_{3}\eta^{3}\cdots,
\end{align}
where%
\begin{equation}
S_{0}=\lim_{\Lambda\rightarrow\infty}\left[  \sum_{\vec{n}\neq\vec{0}}%
\frac{\theta(\Lambda^{2}-\vec{n}^{2})}{\vec{n}^{2}}-4\pi\Lambda\right]  ,
\end{equation}%
\begin{equation}
S_{j}=\sum_{\vec{n}\neq\vec{0}}\frac{1}{\left(  \vec{n}^{2}\right)  ^{j+1}%
}\qquad j\geq1.
\end{equation}
The first few coefficients are
\begin{align}
S_{0}  &  =-8.913631,\quad S_{1}=16.532288,\quad S_{2}=8.401924,\quad
S_{3}=6.945808,\nonumber\\
S_{4}  &  =6.426119,\quad S_{5}=6.202149,\quad S_{6}=6.098184,\quad
S_{7}=6.048263.
\end{align}

L\"{u}scher's formula does not include cutoff effects or the contribution from
coupled higher partial waves for particles with spin. \ However we can neglect
such corrections at asymptotically small momenta.\ \ For small momenta we have
the effective range expansion,%
\begin{equation}
p\cot\delta_{0}(p)\approx-\frac{1}{a_{\text{scatt}}}+\frac{1}{2}r_{0}%
p^{2}+\cdots\text{,}%
\end{equation}
where $a_{\text{scatt}}$ is the scattering length and $r_{0}$ is the effective
range. \ In terms of $\eta$, the energy of the two-body scattering state is%
\begin{equation}
E=\frac{p^{2}}{m}=\frac{\eta}{m}\left(  \frac{2\pi}{L}\right)  ^{2}.
\end{equation}
$S$ is an analytic function of $\eta$ near $\eta=0$, and so we can consider
both $E<0$ and $E>0$. \ We decouple the spin-singlet and spin-triplet contact
interactions by expressing $C$ and $C_{I}$ as a linear combination of
coefficients $C_{^{1}S_{0}}$ and $C_{^{3}S_{1}}$,%
\begin{align}
C  &  =\left(  3C_{^{1}S_{0}}+C_{^{3}S_{1}}\right)  /4,\\
C_{I}  &  =\left(  C_{^{1}S_{0}}-C_{^{3}S_{1}}\right)  /4.
\end{align}
The value of $C_{^{3}S_{1}}$ is tuned to give the physical deuteron binding
energy, $-2.224575(9)$ MeV. \ The value of $C_{^{1}S_{0}}$ is tuned using
L\"{u}scher's formula to give the physical $^{1}S_{0}$ scattering length,
$-23.76(1)$ fm. 

At leading order in the two-nucleon system we expect finite cutoff errors to
scale roughly as $O(\Lambda^{-1})$ or $O(a)$. \ On the lattice we can relate
this cutoff error to the probability that both nucleons occupy the same
lattice site. \ This localized two-nucleon state has a large positive
expectation value for the kinetic energy and a large negative expectation
value for the potential energy. \ Let $E_{2}^{\text{localized}}$ be the
expectation value of the total energy. \ $E_{2}^{\text{localized}}$ need not
be small compared with the cutoff energy $\Lambda^{2}/(2m)$. \ Therefore
transfer matrix elements involving this state may have a significant
dependence on the temporal lattice spacing even for $a_{t}^{-1}$ as large as
the cutoff energy. \ This dependence shows up clearly in the $O(\Lambda^{-1})$
cutoff error, and we see the effect in the following results.

For $a=(100$ MeV$)^{-1}$ and $a_{t}=(70$ MeV$)^{-1},(200$ MeV$)^{-1},(10000$
MeV$)^{-1}$ we set the coefficients $C_{^{3}S_{1}}$ and $C_{^{1}S_{0}}$ using
L\"{u}scher's formula. \ We also use L\"{u}scher's formula to determine the
effective range for the $^{1}S_{0}$ partial wave and both the scattering
length and effective range for the $^{3}S_{1}$ partial wave. \ The results are
shown in Table 1.%
\[%
\genfrac{}{}{0pt}{}{\text{Table 1: \ Coefficients and S-wave parameters for
}a=(100\text{ MeV})^{-1}}{%
\begin{tabular} 
[c]{|c|c|c|c|c|c|}\hline
& $C_{^{3}S_{1}}$ (MeV$^{-2}$) & $C_{^{1}S_{0}}$ (MeV$^{-2}$) & $r_{0}%
^{^{1}S_{0}}$ (fm) & $a_{\text{scatt}}^{^{3}S_{1}}$ (fm) & $r_{0}^{^{3}S_{1}}$
(fm)\\\hline
$a_{t}=(70$ MeV$)^{-1}$ & $-5.714\times10^{-5}$ & $-5.021\times10^{-5}$ &
$-0.179(7)$ & $4.153(5)$ & $-0.48(2)$\\\hline
$a_{t}=(200$ MeV$)^{-1}$ & $-6.706\times10^{-5}$ & $-5.794\times10^{-5}$ &
$0.71(2)$ & $4.522(1)$ & $0.30(2)$\\\hline
$a_{t}=(10000$ MeV$)^{-1}$ & $-7.151\times10^{-5}$ & $-6.126\times10^{-5}$ &
$1.03(2)$ & $4.664(1)$ & $0.53(2)$\\\hline
experiment & $-$ & $-$ & $2.75(5)$ & $5.424(4)$ & $1.759(5)$\\\hline
\end{tabular}
}%
\]

Note that the values for $C_{^{1}S_{0}}$ and $C_{^{3}S_{1}}$ are reasonably close to
the ones found at NLO and NNLO
in the continuum formulation \cite{Epelbaum:2001fm}. 
Also, within the pionless framework
both values should be identical in the Wigner symmetry limit.
The errorbars on the lattice data in Table~1 are error estimates from the least squares
fit. \ Since we work at leading order we do not expect agreement with
the experimental values for the effective ranges. \ However it is interesting
to note that the effective ranges are actually negative for the largest
temporal lattice spacing. \ In the following we explain how this happens.

For some small fixed Euclidean time interval $\Delta t$ consider all
transition amplitudes between two-nucleon states. \ If $a_{t}\ll\Delta t$ then
there are many temporal lattice steps in the time interval $\Delta t$. \ Any
transition amplitude involving states with two nucleons close together is
enhanced to some degree by the negative potential energy of the delta function
potential. \ On the other hand if $a_{t}=\Delta t$ then there is only one
temporal lattice step. \ In this case only the forward matrix element for
incoming and outgoing localized two-nucleon states is enhanced by the delta
function potential. \ This produces a sharp central peak in the two-nucleon
wavefunction where the nucleons overlap and explains the decrease in effective
range. \ By the same reasoning we also expect a smaller value for the deuteron
root-mean-square radius $r_{d}$. \ In Table 2 we show results for $r_{d}$ and
the deuteron quadrupole moment, $Q_{d}$, along with corresponding experimental
values. \ The experimental value quoted for $r_{d}$ is for the point proton
root-mean-square radius.%
\[%
\genfrac{}{}{0pt}{}{\text{Table 2: \ Properties of the deuteron for
}a=(100\text{ MeV})^{-1}}{%
\begin{tabular}
[c]{|c|c|c|}\hline
& $r_{d}$ (fm) & $Q_{d}$ (fm$^{2}$)\\\hline
$a_{t}=(70$ MeV$)^{-1}$ & $1.566(1)$ & $0.144(1)$\\\hline
$a_{t}=(200$ MeV$)^{-1}$ & $1.668(1)$ & $0.171(1)$\\\hline
$a_{t}=(10000$ MeV$)^{-1}$ & $1.736(1)$ & $0.179(1)$\\\hline
experiment & $1.9671(6)$ & $0.2859(3)$\\\hline
\end{tabular}
}%
\]
As expected the root-mean-square radius of the deuteron is smaller than the
physical value, and the deviation is greater for larger values of $a_{t}$.
\ The smaller radius also results in a substantial reduction in the quadrupole moment.

\section{Zero-range clustering instability}

We have found that the deuteron wavefunction at leading order shows some
deficiencies which presumably get fixed at higher order in chiral effective
field theory. \ But since we now have in hand the coefficients of the
leading-order contact interactions for lattice spacing $a=(100$ MeV$)^{-1}$
and $a_{t}=(70$ MeV$)^{-1},(200$ MeV$)^{-1},(10000$ MeV$)^{-1}$, we can consider systems with more than two
nucleons at leading order in chiral effective field theory. \ Unfortunately
here we find more problems. \ In the helium-4 system we discover that the
ground state is severely overbound and consists almost entirely of the quantum
state with all four nucleons occupying the same lattice site. \ This
clustering instability can be understood as the result of two contributing
factors. \ The first is that chiral effective field theory at leading order
gives a poor description of S-wave scattering above a center of mass momentum
of $50$ MeV. \ The leading-order contact interactions are momentum independent
and, as a result, are too strong at high momenta. \ The second is a
combinatorial enhancement of the contact interactions when more than two
nucleons occupy the same lattice site. \ This effect has been studied in
two-dimensional large-$N$ droplets with zero-range attraction
\cite{Lee:2005xy}. \ Similar effects have also been considered in systems of
higher-spin fermions in optical traps and lattices \cite{Wu:2003a,Wu:2004a}.
\ To illustrate we briefly discuss how the problem arises in $SU(4)$-symmetric
pionless theory at leading order using a Hamiltonian lattice formalism.

Let $E_{1}^{\text{localized}}$ be the expectation value for the kinetic energy
of a single nucleon localized on a single lattice site and let $V_{2}<0$ be
the potential energy between two nucleons on the same lattice site. \ If we
fix the two-particle scattering length then both $E_{1}^{\text{localized}}$
and $V_{2}$ scale linearly with the cutoff energy,%
\begin{equation}
E_{1}^{\text{localized}}\sim-V_{2}\sim\frac{\Lambda^{2}}{2m},\qquad\Lambda=\pi
a^{-1}.
\end{equation}
A detailed calculation of $V_{2}$ for infinite scattering length can be found
in \cite{Lee:2005it}. \ The total energies associated with putting two, three,
or four nucleons on the same lattice site are%
\begin{align}
E_{2}^{\text{localized}}  &  =2E_{1}^{\text{localized}}+V_{2},\\
E_{3}^{\text{localized}}  &  =3E_{1}^{\text{localized}}+3V_{2},\\
E_{4}^{\text{localized}}  &  =4E_{1}^{\text{localized}}+6V_{2}.
\end{align}
An instability forms as we increase $A$ because the kinetic energy scales as
the number of nucleons, $A,$ while the potential energy scales as $\binom
{A}{2}$. \ Of course the Pauli exclusion principle prevents more than four
nucleons from sitting on the same lattice site, and so the problem is most
severe in the four-nucleon system.

In the leading-order pionless theory it has been shown that $V_{2}
<-E_{1}^{\text{localized}}$ \cite{Lee:2005it}. \ Therefore $E_{3}
^{\text{localized}}$ is negative and scales with the cutoff energy. \ This
produces an instability for the three-nucleon system in the absence of
three-body forces or other stabilizing effects. \ The instability of the
triton for zero range forces was first studied by Thomas
\cite{Thomas:1935}. \ There have been a number of more recent studies
of the triton in pionless effective field theory as well as more
general three-body systems with short range interactions and long scattering
lengths
\cite{Efimov:1971a,Efimov:1993a,Bedaque:1998kg,Bedaque:1998km,Bedaque:1999ve,Braaten:2004a,Borasoy:2005yc,Platter:2006ad}%
. \ It has also been shown that when the cutoff dependence in the
three-nucleon system is removed using a three-nucleon contact interaction,
then the binding energy of the four-nucleon system appears also to be cutoff
independent \cite{Platter:2004pra,Platter:2004zs}. \ In our lattice
Hamiltonian notation we denote $V_{3}$ as the potential energy associated with
the three-nucleon contact interaction. \ The new localized energies are then%
\begin{align}
E_{2}^{\text{localized}}  &  =2E_{1}^{\text{localized}}+V_{2},\\
E_{3}^{\text{localized}}  &  =3E_{1}^{\text{localized}}+3V_{2}+V_{3},\\
E_{4}^{\text{localized}}  &  =4E_{1}^{\text{localized}}+6V_{2}+4V_{3}.
\end{align}
Clearly $E_{4}^{\text{localized}}$ would be stabilized by $4V_{3}$ for
sufficiently large $V_{3}>0$. \ However for realistic nuclear binding
energies, it was found that the desired cutoff independence in helium-4 does
not emerge until the cutoff momentum $\Lambda$ exceeds $8$ fm$^{-1}$
\cite{Platter:2004zs}. \ Unfortunately this high cutoff momentum makes it a
difficult starting point for lattice simulations of realistic light nuclei.
\ A cutoff momentum of $8$ fm$^{-1}$ corresponds with a lattice spacing of
about $0.4$ fm. \ From a computational standpoint this combination of short
lattice spacing and strong repulsive forces makes lattice simulations nearly
impossible due to sign and phase oscillations. \ Given these difficulties we try a different
approach. \ We keep the lattice spacing large, $a\sim(100$ MeV$)^{-1}\sim2$
fm, and again consider chiral effective field theory at leading order. \ But
this time we introduce higher-derivative operators which improve the S-wave
scattering amplitude at higher momenta. \ We expect that this should remove
the clustering instability in the four-nucleon system.

\section{Higher-derivative terms}  \label{sec:higherderivatives}

\begin{figure}
[bt]
\begin{center}
\includegraphics[
height=3.6288in,
width=3.3857in
]%
{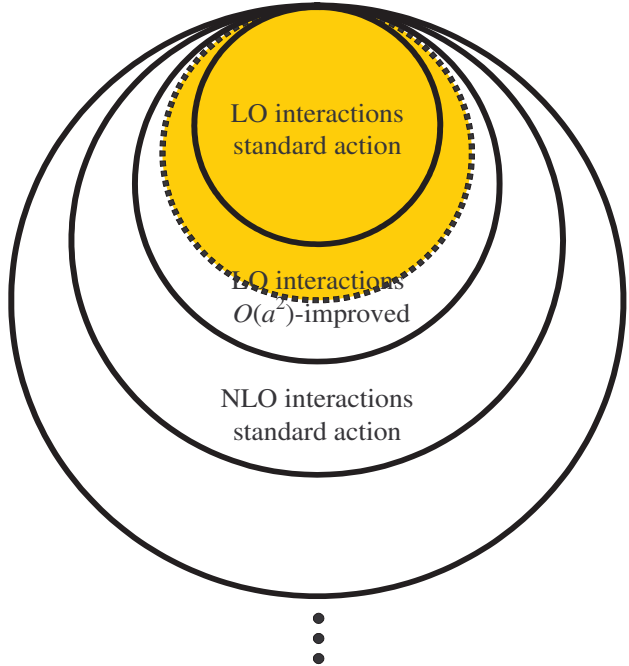}%
\caption{Suppression of finite cutoff errors by broadening the leading order contact interactions.}%
\label{fig:lobroadened}%
\end{center}
\end{figure}

As explained in the previous section, the interactions at leading order with delta function contact
interactions are too strong at large momenta. Their contribution would be appropriately weakened
by interactions of higher chiral order.
But a full investigation of higher order contributions is beyond the scope of this
first exploratory study and is deferred to future work.  Instead we consider here the 
effect of higher derivative terms which reduce cutoff errors by improving the 
delta function contact interaction.
We fix the problem of clustering instability by introducing an ${\cal O}(a^2)$-improved 
broadening for the leading contact interactions $C$ and $C_I$, as illustrated in 
Fig.~\ref{fig:lobroadened}.
This is by no means a full NLO calculation, but rather an LO calculation with ${\cal O}(a^2)$-improvement 
to reduce cutoff errors.

\ To this aim, we define the
momentum-dependent densities,%
\begin{equation}
\rho^{a^{\dag},a}(\vec{q}_{s})=\sum_{\vec{n}_{s}}\rho^{a^{\dag},a}(\vec{n}%
_{s})e^{i\vec{q}_{s}\cdot\vec{n}_{s}},
\end{equation}%
\begin{equation}
\rho_{I}^{a^{\dag},a}(\vec{q}_{s})=\sum_{\vec{n}_{s}}\rho_{I}^{a^{\dag}%
,a}(\vec{n}_{s})e^{i\vec{q}_{s}\cdot\vec{n}_{s}},
\end{equation}
where $\vec{q}_{s}$ is the spatial momentum on the lattice. \ We can write
$\vec{q}_{s}$ as%
\begin{equation}
\vec{q}_{s}=\frac{2\pi}{L}\vec{k}_{s},
\end{equation}
where the components of $\vec{k}_{s}$ are integers from $0$ to $L-1$.

The transfer matrix $M^{(n_{t})}$ with only nucleon fields was defined in
(\ref{transfer_nucleon_only}). \ The contact interactions in $M^{(n_{t})}$
have the form%
\begin{align}
&  -\frac{1}{2}C\alpha_{t}\sum_{\vec{n}_{s}}\left[  \rho^{a^{\dag},a}(\vec
{n}_{s})\right]  ^{2}-\frac{1}{2}C_{I}\alpha_{t}\sum_{I=1,2,3}\sum_{\vec
{n}_{s}}\left[  \rho_{I}^{a^{\dag},a}(\vec{n}_{s})\right]  ^{2}\nonumber\\
&  =\frac{1}{L^{3}}\sum_{\vec{q}_{s}}\left[  -\frac{1}{2}C\alpha_{t}%
\rho^{a^{\dag},a}(\vec{q}_{s})\rho^{a^{\dag},a}(-\vec{q}_{s})-\frac{1}{2}%
C_{I}\alpha_{t}\sum_{I=1,2,3}\rho_{I}^{a^{\dag},a}(\vec{q}_{s})\rho
_{I}^{a^{\dag},a}(-\vec{q}_{s})\right]  .
\end{align}
We replace these by the momentum-dependent interactions,%
\begin{equation}
\frac{1}{L^{3}}\sum_{\vec{q}_{s}}f(\vec{q}_{s}^{\ 2})\left[  -\frac{1}%
{2}C\alpha_{t}\rho^{a^{\dag},a}(\vec{q}_{s})\rho^{a^{\dag},a}(-\vec{q}%
_{s})-\frac{1}{2}C_{I}\alpha_{t}\sum_{I=1,2,3}\rho_{I}^{a^{\dag},a}(\vec
{q}_{s})\rho_{I}^{a^{\dag},a}(-\vec{q}_{s})\right]  ,
\label{momentum_dependent}%
\end{equation}
where the coefficient function $f(\vec{q}_{s}^{\ 2})$ is defined as
\begin{equation}
f(\vec{q}_{s}^{\ 2})=f_{0}^{-1}\exp\left[  -b%
{\displaystyle\sum\limits_{l_{s}=1,2,3}}
\left(  1-\cos q_{l_{s}}\right)  \right]  ,
\end{equation}
and the normalization factor $f_{0}$ is determined by the condition%
\begin{equation}
f_{0}=\frac{1}{L^{3}}\sum_{\vec{q}_{s}}\exp\left[  -b%
{\displaystyle\sum\limits_{l_{s}=1,2,3}}
\left(  1-\cos q_{l_{s}}\right)  \right]  .
\end{equation}
The coefficient $b$ is determined at a later stage when we find the effective
range. \ For small $\vec{q}_{s}$ we see that $f(\vec{q}_{s}^{\ 2})$ reduces to
a Gaussian form,%
\begin{equation}
f(\vec{q}_{s}^{\ 2})\approx f_{0}^{-1}\exp\left(  -\frac{b}{2}\vec{q}%
_{s}^{\ 2}\right)  .
\end{equation}

We can introduce exactly the same momentum-dependent interactions in the
transfer matrix formalism with auxiliary fields,%
\begin{equation}
\mathcal{Z}_{LO}\propto%
{\displaystyle\int}
D\pi_{I}^{\prime}DsDs_{I}\;\exp\left[  -S_{\pi\pi}-S_{ss}\right]  \times
{\rm Tr}\left\{  M^{(L_{t}-1)}(\pi_{I}^{\prime},s,s_{I})\times\cdots\times
M^{(0)}(\pi_{I}^{\prime},s,s_{I})\right\}  .
\end{equation}
To do this we replace%
\begin{equation}
S_{ss}=\frac{1}{2}\sum_{\vec{n}}s^{2}(\vec{n})+\frac{1}{2}\sum_{I}\sum
_{\vec{n}}\left[  s_{I}(\vec{n})\right]  ^{2}%
\end{equation}
by the nonlocal action%
\begin{equation}
\frac{1}{2}\sum_{\vec{n}_{s},\vec{n}_{s}^{\prime},n_{t}}s(\vec{n}_{s}%
,n_{t})f^{-1}(\vec{n}_{s}-\vec{n}_{s}^{\prime})s(\vec{n}_{s}^{\prime}%
,n_{t})+\frac{1}{2}\sum_{I}\sum_{\vec{n}_{s},\vec{n}_{s}^{\prime},n_{t}}%
s_{I}(\vec{n}_{s},n_{t})f^{-1}(\vec{n}_{s}-\vec{n}_{s}^{\prime})s_{I}(\vec
{n}_{s}^{\prime},n_{t}).
\end{equation}
The function $f^{-1}$ is defined as%
\begin{equation}
f^{-1}(\vec{n}_{s}-\vec{n}_{s}^{\prime})=\frac{1}{L^{3}}\sum_{\vec{q}_{s}%
}\frac{1}{f(\vec{q}_{s}^{\ 2})}e^{-i\vec{q}_{s}\cdot(\vec{n}_{s}-\vec{n}%
_{s}^{\prime})}.
\end{equation}
When the auxiliary fields are integrated out we recover the momentum-dependent
interactions in (\ref{momentum_dependent}).

\section{The two-nucleon system revisited} \label{sec:twonucleonrev}

Using the new transfer matrix with momentum-dependent interactions we now
revisit the two-nucleon system. \ Just as before we 
set $C_{^{3}S_{1}}$ and $C_{^{1}S_{0}}$ to give the physical deuteron
binding energy and physical $^{1}S_{0}$ scattering length. \ We also tune the
coefficient $b$ so that when $C_{^{3}S_{1}}$ and $C_{^{1}S_{0}}$ are
determined, we also get the correct value for the average effective range
$\frac{1}{2}\left(  r_{0}^{^{1}S_{0}}+r_{0}^{^{3}S_{1}}\right)  .$ \ For 
$a = (100 \ \mbox{MeV})^{-1}$ and $a_{t}= (70 \ \mbox{MeV})^{-1}$ we find
$C_{^{3}S_{1}}=-4.780\times10^{-5}$ MeV$^{-2}$, $C_{^{1}S_{0}}=-3.414\times
10^{-5}$ MeV$^{-2}$, and $b=0.6$. \ The new results are shown in Tables 3 and
4.%
\[%
\genfrac{}{}{0pt}{}{\text{Table 3: S-wave parameters}}{%
\begin{tabular}
[c]{|c|c|c|c|}\hline
& $r_{0}^{^{1}S_{0}}$ (fm) & $a_{\text{scatt}}^{^{3}S_{1}}$ (fm) &
$r_{0}^{^{3}S_{1}}$ (fm)\\\hline
lattice & $3.20(1)$ & $5.30(1)$ & $1.46(3)$\\\hline
experiment & $2.75(5)$ & $5.424(4)$ & $1.759(5)$\\\hline
\end{tabular}
}%
\]%
\[%
\genfrac{}{}{0pt}{}{\text{Table 4: \ Properties of the deuteron}}{%
\begin{tabular}
[c]{|c|c|c|}\hline
& $r_{d}$ (fm) & $Q_{d}$ (fm$^{2}$)\\\hline
lattice & $1.989(1)$ & $0.278(1)$\\\hline
experiment & $1.9671(6)$ & $0.2859(3)$\\\hline
\end{tabular}
}%
\]
The agreement with experimental values is now good. \ There is clear
improvement over the earlier results shown in Tables 1 and 2.

\begin{figure}
[ptb]
\begin{center}
\includegraphics[
height=3.6288in,
width=3.3857in
]%
{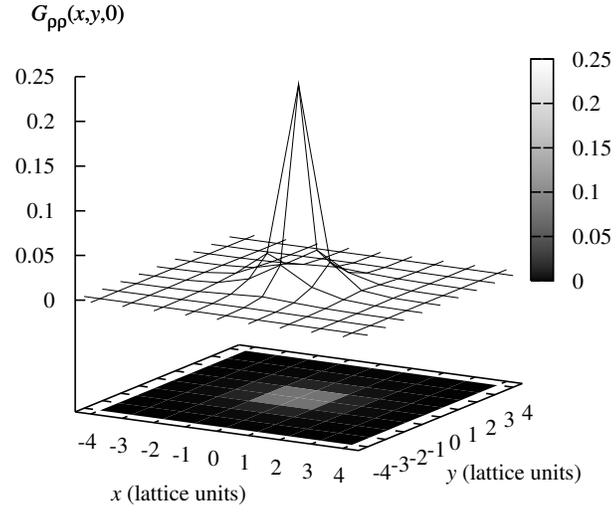}%
\caption{Density correlations in the $xy$-plane for a deuteron with spin in
the $+z$-direction.}%
\label{xyplane_sz1}%
\end{center}
\end{figure}

\begin{figure}
[ptbptb]
\begin{center}
\includegraphics[
height=3.6288in,
width=3.3857in
]%
{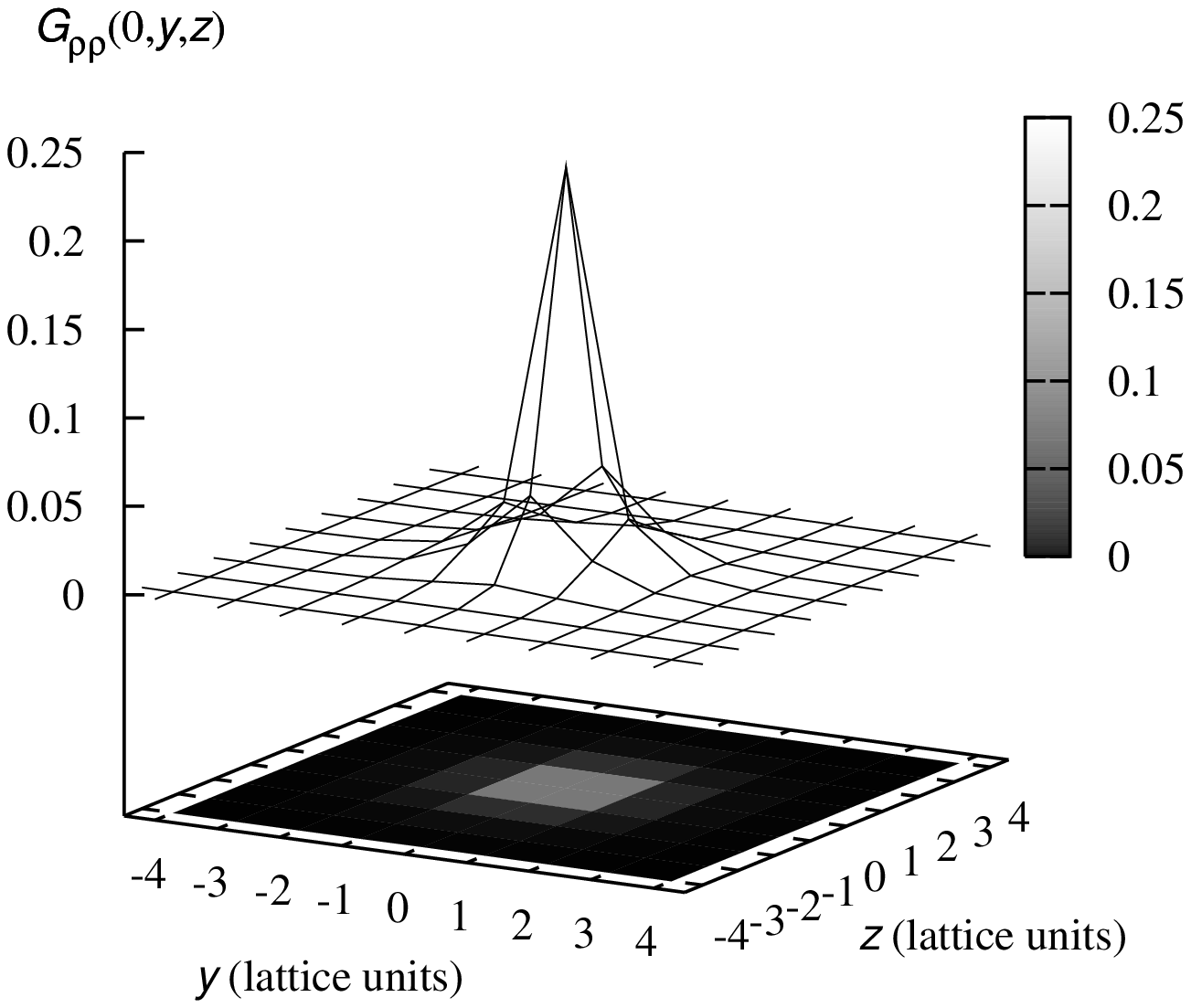}%
\caption{Density correlations in the $yz$-plane for a deuteron with spin in
the $+z$-direction.}%
\label{yzplane_sz1}%
\end{center}
\end{figure}

\begin{figure}
[ptbptbptb]
\begin{center}
\includegraphics[
height=3.6288in,
width=3.3849in
]%
{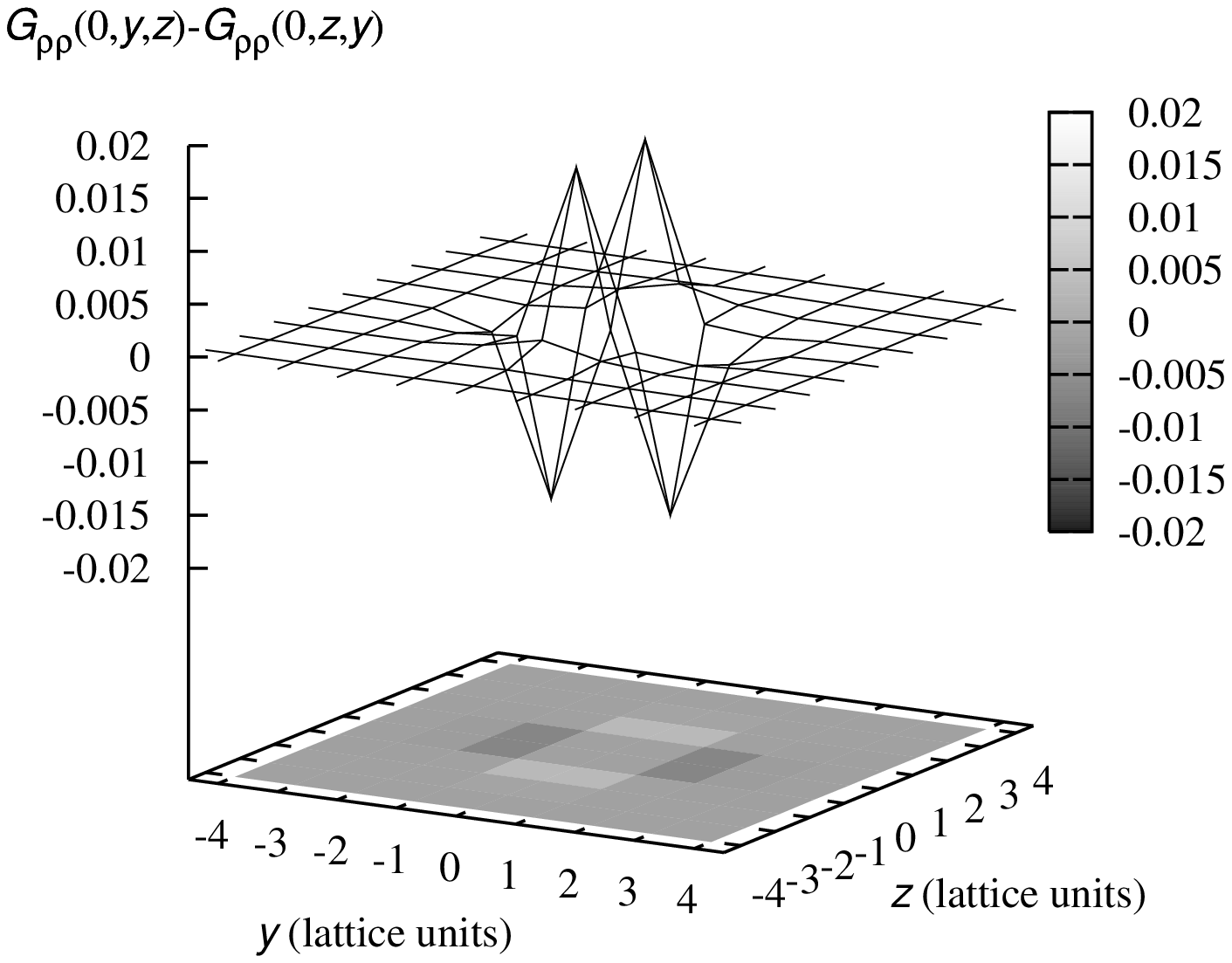}%
\caption{A linear combination of density correlations in the $yz$-plane that
is antisymmetric under interchange of $y$ and $z$. \ The deuteron spin points
in the $+z$-direction.}%
\label{yzplane_sz1_diff}%
\end{center}
\end{figure}

We can probe the shape of deuteron wavefunction by computing the nucleon
density correlation function%
\begin{equation}
\left\langle :\rho^{a^{\dag},a}(\vec{n}_{s})\rho^{a^{\dag},a}(\vec
{0}):\right\rangle .
\end{equation}
If $A$ is the total number of nucleons then
\begin{equation}
A=\sum_{\vec{n}_{s}}\left\langle \rho^{a^{\dag},a}(\vec{n}_{s})\right\rangle .
\end{equation}
We also find%
\begin{align}
\sum_{\vec{n}_{s}}\left\langle :\rho^{a^{\dag},a}(\vec{n}_{s})\rho^{a^{\dag
},a}(\vec{0}):\right\rangle  &  =\sum_{\vec{n}_{s}}\left\langle \rho^{a^{\dag
},a}(\vec{n}_{s})\rho^{a^{\dag},a}(\vec{0})\right\rangle -\left\langle
\rho^{a^{\dag},a}(\vec{0})\right\rangle \nonumber\\
&  =L^{-3}\left(  A^{2}-A\right)  .
\end{align}
Let us define the normalized density correlation function as%
\begin{equation}
G_{\rho\rho}(\vec{n}_{s})=L^{3}\left(  A^{2}-A\right)  ^{-1}\left\langle
:\rho^{a^{\dag},a}(\vec{n}_{s})\rho^{a^{\dag},a}(\vec{0}):\right\rangle .
\label{density_correlation}%
\end{equation}
In Figure~\ref{xyplane_sz1}, we show $G_{\rho\rho}(\vec{n}_{s})$ for the
deuteron in the $xy$-plane. \ We have aligned the deuteron spin in the
$+z$-direction. \ Keeping the deuteron spin in the $+z$-direction, in Figure~\ref{yzplane_sz1} 
we show $G_{\rho\rho}(\vec{n}_{s})$ in the $yz$-plane. \ A
small asymmetry can be seen between the $y$ and $z$ directions. \ This is a
signal of the deuteron quadrupole moment and can be seen more easily in
Fig.~\ref{yzplane_sz1_diff}, where we have taken an antisymmetric combination
under interchange of $y$ and $z$ .%

\section{Transfer matrix projection method for light nuclei}

We simulate light nuclei by using the Monte Carlo transfer matrix projection
method introduced in \cite{Lee:2005fk}. \ Since this method may be unfamiliar,
we first give an overview of the calculation using continuum notation before
describing the details of the lattice transfer matrix calculation.

Let $\left\vert \Psi_{Z,N}^{\text{free}}\right\rangle $ be a Slater
determinant of free particle standing waves in a periodic cube for $Z$ protons
and $N$ neutrons. \ We define $A=Z+N$ as the total number of nucleons. \ Let
$H_{LO}$ denote the Hamiltonian including instantaneous one-pion exchange and
the improved higher-derivative contact interactions. \ Let $H_{SU(4)\not \pi
}$ be the same Hamiltonian but with both $C_{I}$ and $g_{A}$ set to zero. \ As
the notation suggests, $H_{SU(4)\not \pi }$ is invariant under $SU(4)$ Wigner
symmetry. \ Wigner symmetry refers to an idealized limit where spin and
isospin degrees of freedom can be interchanged and the $SU(2)\times SU(2)$
spin-isospin symmetry is elevated to an $SU(4)$ symmetry. \ Let us define a
trial wavefunction%
\begin{equation}
\left\vert \Psi(t^{\prime})\right\rangle =\exp\left[  -H_{SU(4)\not \pi
}t^{\prime}\right]  \left\vert \Psi_{Z,N}^{\text{free}}\right\rangle .
\end{equation}
With this trial wavefunction we define the amplitude,%
\begin{equation}
Z(t)=\left\langle \Psi(t^{\prime})\right\vert \exp\left[  -H_{LO}t\right]
\left\vert \Psi(t^{\prime})\right\rangle ,
\end{equation}
as well as the transient energy,%
\begin{equation}
E(t)=-\frac{\partial}{\partial t}\left[  \ln Z(t)\right]  .
\end{equation}
In limit of large $t$ we get%
\begin{equation}
\lim_{t\rightarrow\infty}E(t)=E_{0},
\end{equation}
where $E_{0}$ is the energy of the lowest eigenstate $\left\vert \Psi
_{0}\right\rangle $ of $H_{LO}$ with a nonzero inner product with $\left\vert
\Psi(t^{\prime})\right\rangle $. \ In order to compute the expectation value
of some normal-ordered operator $O$ we define%
\begin{equation}
Z_{O}(t)=\left\langle \Psi(t^{\prime})\right\vert \exp\left[  -H_{LO}%
t/2\right]  \,O\,\exp\left[  -H_{LO}t/2\right]  \left\vert \Psi(t^{\prime
})\right\rangle .
\end{equation}
The expectation value of $O$ for $\left\vert \Psi_{0}\right\rangle $ can be
computed in the large $t$ limit,%
\begin{equation}
\lim_{t\rightarrow\infty}\frac{Z_{O}(t)}{Z(t)}=\left\langle \Psi
_{0}\right\vert O\left\vert \Psi_{0}\right\rangle .
\end{equation}
In this two-step approach we use $\exp\left[  -H_{SU(4)\not \pi }t^{\prime
}\right]  $ as an approximate inexpensive low-energy filter and $\exp\left[
-H_{LO}t\right]  $ as an exact low-energy filter. \ The projection
$\exp\left[  -H_{SU(4)\not \pi }t^{\prime}\right]  $ is computationally
inexpensive because the path integral for leading-order pionless effective
field theory in the $SU(4)$ limit is strictly positive for any even number of
nucleons \cite{Lee:2004hc,Chen:2004rq}. \ Although there is no positivity
theorem for odd numbers of nucleons, sign oscillations are also suppressed in
odd systems because it is only one particle or one hole away from an even
system with no sign oscillations.

\begin{figure}
[ptb]
\begin{center}
\includegraphics[
height=1.6786in,
width=5.1378in
]%
{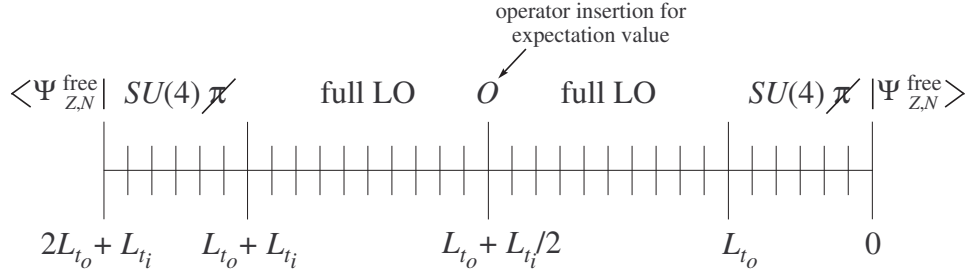}%
\caption{Overview of the various pieces of the transfer matrix calculation.}%
\label{time_steps}%
\end{center}
\end{figure}

In the lattice transfer matrix formalism we construct $\left\vert
\Psi(t^{\prime})\right\rangle $ using%
\begin{equation}
\left\vert \Psi(t^{\prime})\right\rangle =%
{\displaystyle\int}
Ds\;\exp\left[  -S_{ss}(s)\right]  \times M_{SU(4)\not \pi }^{(L_{t_{o}}%
-1)}(s)\times\cdots\times M_{SU(4)\not \pi }^{(0)}(s)\left\vert \Psi
_{Z,N}^{\text{free}}\right\rangle ,
\end{equation}
where $t^{\prime}=L_{t_{o}}a_{t}$. \ $M_{SU(4)\not \pi }^{(i)}(s)$ is the same
as $M^{(i)}(\pi_{I}^{\prime},s,s_{I})$ except with $g_{A}=C_{I}=0$. \ We have
omitted the dependence on $\pi_{I}^{\prime}$ and $s_{I}$ since these
completely decouple in the $SU(4)$ symmetric theory. \ The amplitude $Z(t)$ is
constructed using%
\begin{equation}
Z(t)=%
{\displaystyle\int}
D\pi_{I}^{\prime}DsDs_{I}\;\exp\left[  -S_{\pi\pi}-S_{ss}\right]
\times\left\langle \Psi(t^{\prime})\right\vert M^{(L_{t_{i}}-1)}(\pi
_{I}^{\prime},s,s_{I})\times\cdots\times M^{(0)}(\pi_{I}^{\prime}%
,s,s_{I})\left\vert \Psi(t^{\prime})\right\rangle ,
\end{equation}
where $t=L_{t_{i}}a_{t}$. \ In this case the full leading-order transfer
matrix is used. \ We compute $Z_{O}(t)$ by inserting the normal-ordered
operator $O$ in the middle of the string of transfer matrices,%
\begin{align}
Z_{O}(t)  &  =%
{\displaystyle\int}
D\pi_{I}^{\prime}DsDs_{I}\;\exp\left[  -S_{\pi\pi}-S_{ss}\right]
\times\left\langle \Psi(t^{\prime})\right\vert M^{(L_{t_{i}}-1)}(\pi
_{I}^{\prime},s,s_{I})\times\cdots\nonumber\\
&  \qquad\qquad\cdots\times M^{(L_{t_{i}}/2)}(\pi_{I}^{\prime},s,s_{I}%
)\,O\,M^{(L_{t_{i}}/2-1)}(\pi_{I}^{\prime},s,s_{I})\times\cdots\times
M^{(0)}(\pi_{I}^{\prime},s,s_{I})\left\vert \Psi(t^{\prime})\right\rangle .
\end{align}
A schematic overview of the transfer matrix calculation is shown in Fig.
\ref{time_steps}.%

Let $\left\vert \psi_{1}\right\rangle ,\cdots,\left\vert \psi_{A}\right\rangle
$ be the free particle standing waves comprising $\left\vert \Psi
_{Z,N}^{\text{free}}\right\rangle .$ \ We note that for the transfer matrices
in the auxiliary field formalism there are no direct interactions between
particles, only single-nucleon operators interacting with the background pion
and auxiliary fields. \ It is easier to see this fact if we pretend for the
moment that each of the $A$ nucleons has an extra quantum number which makes
them distinguishable. \ Let us label the extra quantum number as $X$, where
$X=1,\cdots,A.$ \ Then we have%
\begin{equation}
a_{i,j}^{\dagger},a_{i,j}\rightarrow\left\{  a_{i,j,X}^{\dagger}%
,a_{i,j,X}\right\}  _{X=1,\cdots,A}.
\end{equation}
We use an $X$ subscript to indicate this replacement for creation and
annihilation operators. \ The transfer matrices $M_{SU(4)\not \pi }^{(n_{t})}$
and $M^{(n_{t})}$ factorize into transfer matrices for each $X$,%
\begin{align}
M_{SU(4)\not \pi }^{(n_{t})}  &  \rightarrow%
{\displaystyle\prod\limits_{X=1,\cdots,A}}
M_{SU(4)\not \pi ,X}^{(n_{t})},\\
M^{(n_{t})}  &  \rightarrow%
{\displaystyle\prod\limits_{X=1,\cdots,A}}
M_{X}^{(n_{t})}.
\end{align}
\ So long as the initial and final state wavefunctions are completely
antisymmetric in $X$, this error in quantum statistics has no effect on the
final amplitude. \ We can write the full $A$-nucleon transfer matrix element
as a determinant of an $A\times A$ matrix of single-particle matrix elements,%
\begin{align}
\mathcal{M}_{ij}(\pi_{I}^{\prime},s,s_{I},t^{\prime},t)  &  =\left\langle
\psi_{i,X}\right\vert M_{SU(4)\not \pi ,X}^{(2L_{t_{o}}+L_{t_{i}}-1)}%
\times\cdots\times M_{SU(4)\not \pi ,X}^{(L_{t_{o}}+L_{t_{i}})}M_{X}%
^{(L_{t_{o}}+L_{t_{i}}-1)}\times\cdots\nonumber\\
&  \cdots\times M_{X}^{(L_{t_{o}})}M_{SU(4)\not \pi ,X}^{(L_{t_{o}}-1)}%
\times\cdots\times M_{SU(4)\not \pi ,X}^{(0)}\left\vert \psi_{j,X}%
\right\rangle .
\end{align}
The indices $i,j$ go from $1$ to $A$. \ The calculation of $Z(t)$ now reduces
to computing%
\begin{equation}
Z(t)=%
{\displaystyle\int}
D\pi_{I}^{\prime}DsDs_{I}\;\exp\left\{  -S_{\pi\pi}-S_{ss}\right\}
\det\mathcal{M}(\pi_{I}^{\prime},s,s_{I},t^{\prime},t).
\end{equation}

There are many different ways to compute the amplitude $Z_{O}(t)$, and the
most efficient method will depend on the operator $O$. \ Insertions of an
operator measuring spatial difermion pair correlations was discussed in
\cite{Lee:2006hr}. Here we consider an operator that measures spatial
correlations of the total nucleon density,%
\begin{equation}
O= \ :\rho^{a^{\dag},a}(\vec{n}_{s})\rho^{a^{\dag},a}(\vec{0}):.
\end{equation}
For this operator it is convenient to use the identity%
\begin{equation}
O=\lim_{\epsilon_{1},\epsilon_{2}\rightarrow0}\frac{\partial}{\partial
\epsilon_{1}}\frac{\partial}{\partial\epsilon_{2}}M(\epsilon_{1},\epsilon
_{2}),
\end{equation}
where%
\begin{equation}
M(\epsilon_{1},\epsilon_{2})= \ :\exp\left[  \epsilon_{1}\sum_{i,j}a_{i,j}^{\dag
}(\vec{n}_{s})a_{i,j}(\vec{n}_{s})+\epsilon_{2}\sum_{i,j}a_{i,j}^{\dag}%
(\vec{0})a_{i,j}(\vec{0})\right]  :.
\end{equation}
This is useful because $M(\epsilon_{1},\epsilon_{2})$ itself looks like a
transfer matrix with only single-nucleon operators. \ Let us define the new
single-particle matrix elements,%
\begin{align}
&  \mathcal{M}_{ij}(\pi_{I}^{\prime},s,s_{I},t^{\prime},t,\epsilon
_{1},\epsilon_{2})\nonumber\\
&  =\left\langle \psi_{i,X}\right\vert M_{SU(4)\not \pi ,X}^{(2L_{t_{o}%
}+L_{t_{i}}-1)}\times\cdots\times M_{SU(4)\not \pi ,X}^{(L_{t_{o}}+L_{t_{i}}%
)}M_{X}^{(L_{t_{o}}+L_{t_{i}}-1)}\times\cdots\nonumber\\
&  \cdots\times M_{X}^{(L_{t_{o}}+L_{t_{i}}/2)}M_{X}(\epsilon_{1},\epsilon
_{2})M_{X}^{(L_{t_{o}}+L_{t_{i}}/2-1)}\times\cdots\times M_{X}^{(L_{t_{o}}%
)}M_{SU(4)\not \pi ,X}^{(L_{t_{o}}-1)}\times\cdots\times M_{SU(4)\not \pi
,X}^{(0)}\left\vert \psi_{j,X}\right\rangle .
\end{align}
We then find%
\begin{equation}
Z_{O}(t)=\lim_{\epsilon_{1},\epsilon_{2}\rightarrow0}\frac{\partial}%
{\partial\epsilon_{1}}\frac{\partial}{\partial\epsilon_{2}}%
{\displaystyle\int}
D\pi_{I}^{\prime}DsDs_{I}\;\exp\left\{  -S_{\pi\pi}-S_{ss}\right\}
\det\mathcal{M}(\pi_{I}^{\prime},s,s_{I},t^{\prime},t,\epsilon_{1}%
,\epsilon_{2}).
\end{equation}

We use hybrid Monte Carlo to update the fields $\pi_{I}^{\prime},s,s_{I}$
\cite{Duane:1987de}. \ We introduce conjugate fields $p_{\pi_{I}^{\prime}%
},p_{s},p_{s_{I}}$ and use molecular dynamics trajectories to generate new
configurations for $p_{\pi_{I}^{\prime}},p_{s},$ $p_{s_{I}},\pi_{I}^{\prime
},s,s_{I}$ which keep%
\begin{equation}
H_{HMC}=\frac{1}{2}\sum_{I,\vec{n}}p_{\pi_{I}^{\prime}}^{2}(\vec{n})+\frac
{1}{2}\sum_{\vec{n}}p_{s}^{2}(\vec{n})+\frac{1}{2}\sum_{I,\vec{n}}p_{s_{I}%
}^{2}(\vec{n})+V(\pi_{I}^{\prime},s,s_{I}) \label{H_HMC}%
\end{equation}
constant, where%
\begin{equation}
V(\pi_{I}^{\prime},s,s_{I})=S_{\pi\pi}+S_{ss}-\log\left\{  \left\vert \det
\mathcal{M}(\pi_{I}^{\prime},s,s_{I},t^{\prime},t_{\text{end}})\right\vert \right\}  .
\label{V}%
\end{equation}
$t_{\text{end}}$ denotes the largest value of $t$ being considered. \ Upon
completion of each molecular dynamics trajectory, we apply a Metropolis accept
or reject step for the new configuration according to the probability
distribution $e^{-H_{HMC}}$. \ This process of molecular dynamics trajectory
and Metropolis step is repeated many times. \ Each time for the current
configuration$,\mathcal{C}$, we measure%
\begin{equation}
e^{i\theta}(\mathcal{C})=\frac{\det\mathcal{M}(\pi_{I}^{\prime},s,s_{I},t^{\prime
},t_{\text{end}})}{\left\vert \det\mathcal{M}(\pi_{I}^{\prime},s,s_{I},t^{\prime
},t_{\text{end}})\right\vert }, \label{e_i_theta}%
\end{equation}%
\begin{equation}
Z(t,\mathcal{C})=\frac{\det\mathcal{M}(\pi_{I}^{\prime},s,s_{I},t^{\prime}%
,t)}{\left\vert \det\mathcal{M}(\pi_{I}^{\prime},s,s_{I},t^{\prime},t_{\text{end}%
})\right\vert },
\end{equation}
and%
\begin{equation}
Z_{O}(\epsilon_{1},\epsilon_{2},\mathcal{C})=\frac{\det\mathcal{M}(\pi
_{I}^{\prime},s,s_{I},t^{\prime},t_{\text{end}},\epsilon_{1},\epsilon_{2}%
)}{\left\vert \det\mathcal{M}(\pi_{I}^{\prime},s,s_{I},t^{\prime},t_{\text{end}%
})\right\vert }.
\end{equation}
We take the ensemble averages for each measurement and form the ratios%
\begin{equation}
\frac{Z(t)}{Z(t_{\text{end}})}=\frac{\left\langle Z(t,\mathcal{C}%
)\right\rangle }{\left\langle e^{i\theta}(\mathcal{C})\right\rangle },
\end{equation}
and%
\begin{equation}
\frac{Z_{O}(t_{\text{end}})}{Z(t_{\text{end}})}=\lim_{\epsilon_{1}%
,\epsilon_{2}\rightarrow0}\frac{\partial}{\partial\epsilon_{1}}\frac{\partial
}{\partial\epsilon_{2}}\frac{\left\langle Z_{O}(\epsilon_{1},\epsilon
_{2},\mathcal{C})\right\rangle }{\left\langle e^{i\theta}(\mathcal{C}%
)\right\rangle }.
\end{equation}
The ground state energy $E_{0}$ is extracted using%
\begin{equation}
\lim_{t_{\text{end}}\rightarrow\infty}\frac{Z(t_{\text{end}}-\Delta
t)}{Z(t_{\text{end}})}=\exp\left(  E_{0}\Delta t\right)  ,
\end{equation}
and the expectation value $\left\langle \Psi_{0}\right\vert O\left\vert
\Psi_{0}\right\rangle $ is calculated using%
\begin{equation}
\lim_{t_{\text{end}}\rightarrow\infty}\frac{Z_{O}(t_{\text{end}}%
)}{Z(t_{\text{end}})}=\left\langle \Psi_{0}\right\vert O\left\vert \Psi
_{0}\right\rangle .
\end{equation}

\section{Numerical checks using the two-nucleon system}

We have already calculated properties of the deuteron using the transfer
matrix with pions and auxiliary fields integrated out. \ In this section we
make use of the two-nucleon system as a high-precision test of the transfer
matrix Monte Carlo code. \ We calculate the same observables using both the
Monte Carlo code and the exact transfer matrix, 
which is referred to as ``Exact'' method in the following. \ We use the lattice
parameters defined previously, $\ a=(100$ MeV$)^{-1}$, $a_{t}=(70$ MeV$)^{-1}%
$, $C_{^{3}S_{1}}=-4.780\times10^{-5}$ MeV$^{-2}$, $C_{^{1}S_{0}}%
=-3.414\times10^{-5}$ MeV$^{-2}$, and $b=0.6$. \ We set $L=3$, $L_{t_{o}}=2,$
$L_{t_{i}}=2$. \ We consider a small system so that the stochastic error is
small enough to detect disagreement at the $0.1\%-1\%$ level.

For the first test we consider $\left\vert \Psi_{Z,N}^{\text{free}%
}\right\rangle $ with $Z=0,N=2$. \ The standing waves $\psi_{1,2}$ comprising $\left\vert
\Psi_{Z,N}^{\text{free}}\right\rangle $ are%
\begin{align}
\left\langle 0\right\vert a_{i,j}(\vec{n}_{s})\left\vert \psi_{1}%
\right\rangle  &  =L^{-3/2}\delta_{i,0}\delta_{j,1},\nonumber\\
\left\langle 0\right\vert a_{i,j}(\vec{n}_{s})\left\vert \psi_{2}%
\right\rangle  &  =L^{-3/2}\delta_{i,1}\delta_{j,1}.
\end{align}
This corresponds with a $J_{z}=0$, $J=0$ dineutron system with zero total
momentum. \ We compute $E(t)$ as well as the density correlation%
\begin{equation}
G_{\rho\rho}(\vec{n}_{s})=L^{3}\left(  A^{2}-A\right)  ^{-1}\left\langle
:\rho^{a^{\dag},a}(\vec{n}_{s})\rho^{a^{\dag},a}(\vec{0}):\right\rangle .
\end{equation}
From $G_{\rho\rho}(\vec{n}_{s})$ we can determine the root-mean-square radius
$r_{\text{RMS}}$ as well as the quadrupole moment $Q$. \ The results for
$G_{\rho\rho}(\vec{n}_{s})$ are shown in Table 5 and the results for $E(t)$,
$r_{\text{RMS}},$ and $Q$ are shown in Table 6.%

\[%
\genfrac{}{}{0pt}{}{\text{Table 5: \ }G_{\rho\rho}(\vec{n}_{s})\text{ for the
}J_{z}=0,J=0\text{ dineutron system}}{%
\begin{tabular}
[c]{||c||c|c||}\hline
$\vec{n}_{s}$ & Monte Carlo & Exact\\\hline
$(0,0,0)$ & $0.0960(3)$ & $0.09575$\\\hline
$(1,0,0)$ & $0.04506(5)$ & $0.04508$\\\hline
$(0,1,0)$ & $0.04515(6)$ & $0.04508$\\\hline
$(0,0,1)$ & $0.04500(5)$ & $0.04508$\\\hline
$(0,1,1)$ & $0.03354(4)$ & $0.03363$\\\hline
$(1,0,1)$ & $0.03365(4)$ & $0.03363$\\\hline
$(1,1,0)$ & $0.03358(6)$ & $0.03363$\\\hline
$(1,1,1)$ & $0.02876(4)$ & $0.02878$\\\hline
\end{tabular}
\ }%
\]

\[%
\genfrac{}{}{0pt}{}{\text{Table 6: \ }E(t)\text{, }r_{\text{RMS}}\text{, and
}Q\text{ for the }J_{z}=0,J=0\text{ dineutron system}}{%
\begin{tabular}
[c]{||c||c|c||}\hline
& Monte Carlo & Exact\\\hline
$E(t)$ (MeV) & $-5.84(9)$ & $-5.917$\\\hline
$r_{\text{RMS}}$ (fm) & $1.402(4)$ & $1.4015$\\\hline
$\left\vert Q\right\vert $ (fm$^{2}$) & $<10^{-5}$ & $0$\\\hline
\end{tabular}
\ }%
\]
The final values are not of physical relevance since the volume and number of
time steps are very small. \ The\ important result is we find no disagreement
between Monte Carlo results and the exact results beyond the stochastic error level.

For the second test we consider $\left\vert \Psi_{Z,N}^{\text{free}%
}\right\rangle $ with $Z=1,N=1$. \ This time the standing waves comprising
$\left\vert \Psi_{Z,N}^{\text{free}}\right\rangle $ are%
\begin{align}
\left\langle 0\right\vert a_{i,j}(\vec{n}_{s})\left\vert \psi_{1}%
\right\rangle  &  =L^{-3/2}\delta_{i,0}\delta_{j,1},\nonumber\\
\left\langle 0\right\vert a_{i,j}(\vec{n}_{s})\left\vert \psi_{2}%
\right\rangle  &  =L^{-3/2}\delta_{i,0}\delta_{j,0}.
\end{align}
This corresponds with a deuteron system with $J_{z}=1$, $J=1$ and zero total
momentum. \ The results for $G_{\rho\rho}(\vec{n}_{s})$ are shown in Table 7
and the results for $E(t)$, $r_{\text{RMS}},$ and $Q$ are shown in Table 8.%

\[%
\genfrac{}{}{0pt}{}{\text{Table 7: \ }G_{\rho\rho}(\vec{n}_{s})\text{ for the
}J_{z}=1,J=1\text{ deuteron system}}{%
\begin{tabular}
[c]{||c||c|c||}\hline
$(n_{x},n_{y},n_{z})$ & Monte Carlo & Exact\\\hline
$(0,0,0)$ & $0.1230(3)$ & $0.12262$\\\hline
$(1,0,0)$ & $0.04233(4)$ & $0.04240$\\\hline
$(0,1,0)$ & $0.04247(5)$ & $0.04240$\\\hline
$(0,0,1)$ & $0.05563(5)$ & $0.05572$\\\hline
$(0,1,1)$ & $0.03415(5)$ & $0.03422$\\\hline
$(1,0,1)$ & $0.03423(3)$ & $0.03422$\\\hline
$(1,1,0)$ & $0.02764(4)$ & $0.02766$\\\hline
$(1,1,1)$ & $0.02650(3)$ & $0.02649$\\\hline
\end{tabular}
\ }%
\]%
\[%
\genfrac{}{}{0pt}{}{\text{Table 8: \ }E(t)\text{, }r_{\text{RMS}}\text{, and
}Q\text{ for the }J_{z}=1,J=1\text{ deuteron system}}{%
\begin{tabular}
[c]{||c||c|c||}\hline
& Monte Carlo & Exact\\\hline
$E(t)$ (MeV) & $-9.26(9)$ & $-9.311$\\\hline
$r_{\text{RMS}}$ (fm) & $0.6957(2)$ & $0.69564$\\\hline
$Q$ (fm$^{2}$) & $0.1026(2)$ & $0.10283$\\\hline
\end{tabular}
\ }%
\]
Again the final values are not of physical relevance since the volume and
number of time steps are small. \ We find no disagreement between Monte Carlo
results and the exact results beyond the stochastic error level.

\section{Results for the triton}

In our calculations we assume isospin symmetry and so our results for helium-3
and the triton will be the same. \ We choose to compare with experimental data
for the triton since it is not affected by electrostatic repulsion. \ For our
simulations we use the lattice parameters$\ a=(100$ MeV$)^{-1}$, $a_{t}=(70$
MeV$)^{-1}$, $C_{^{3}S_{1}}=-4.780\times10^{-5}$ MeV$^{-2}$, $C_{^{1}S_{0}%
}=-3.414\times10^{-5}$ MeV$^{-2}$, and $b=0.6$. \ We set $L=5$, $L_{t_{o}}=8$
and vary $L_{t_{i}}$ from $2$ to $10$. The standing waves comprising
$\left\vert \Psi_{Z,N}^{\text{free}}\right\rangle $ are%
\begin{align}
\left\langle 0\right\vert a_{i,j}(\vec{n}_{s})\left\vert \psi_{1}%
\right\rangle  &  =L^{-3/2}\delta_{i,0}\delta_{j,1},\nonumber\\
\left\langle 0\right\vert a_{i,j}(\vec{n}_{s})\left\vert \psi_{2}%
\right\rangle  &  =L^{-3/2}\delta_{i,0}\delta_{j,0},\nonumber\\
\left\langle 0\right\vert a_{i,j}(\vec{n}_{s})\left\vert \psi_{3}%
\right\rangle  &  =L^{-3/2}\delta_{i,1}\delta_{j,1}.
\end{align}
This corresponds to a triton system with $J_{z}=\frac{1}{2},J=\frac{1}{2}$
and total momentum zero.

\begin{figure}
[ptb]
\begin{center}
\includegraphics[
height=3.026in,
width=3.3875in
]%
{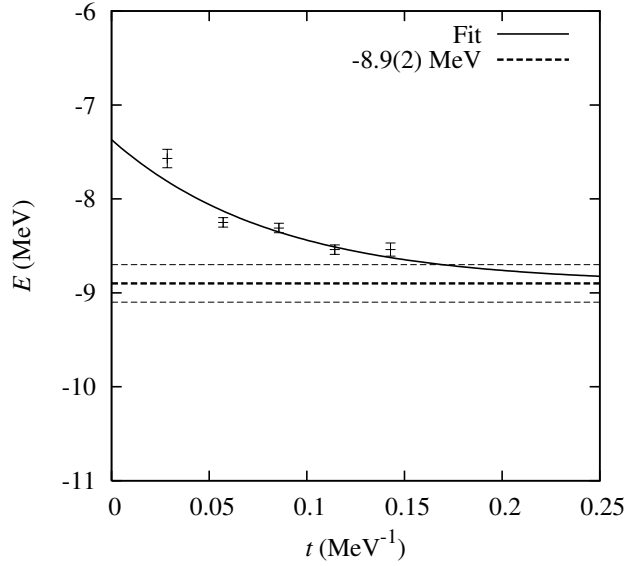}%
\caption{The transient energy for the triton system as a function of imaginary
time $t$.}%
\label{triton_energy}%
\end{center}
\end{figure}

For each value of $L_{t_{i}}$ a total of about $5\times10^{6}$ hybrid Monte
Carlo trajectories are generated by $2048$ processors, each running completely
independent trajectories. \ Averages and stochastic errors are computed by
comparing the results of all $2048$ processors. \ In Fig. \ref{triton_energy}
we show the triton energy as a function of imaginary time $t$. \ The error
bars denote the stochastic error. \ To remove the transient signal of higher
energy states, we fit $E(t)$ to a decaying exponential form at large $t$,%
\begin{equation}
E(t)\rightarrow E_{0}+ce^{-\Delta Et}.
\end{equation}
A least squares fit gives an asymptotic value $E_{0}=-8.9(2)$\ MeV. \ Our
result is within $5\%$ agreement with the experimental value of $-8.48$ MeV.%
\ Note that the triton appears to be overbound. However, for the triton
the lattice volume of $(9.85 \ \mbox{fm})^3$ that we use might still be too small and
finite volume effects can possibly occur. In this case increasing the lattice volume
would lead to a slightly less bound triton. Both finite volume effects
and the systematic inclusion of higher chiral orders
in the effective potentials will be investigated in future work.

\begin{figure}
[ptb]
\begin{center}
\includegraphics[
height=3.6288in,
width=3.3857in
]%
{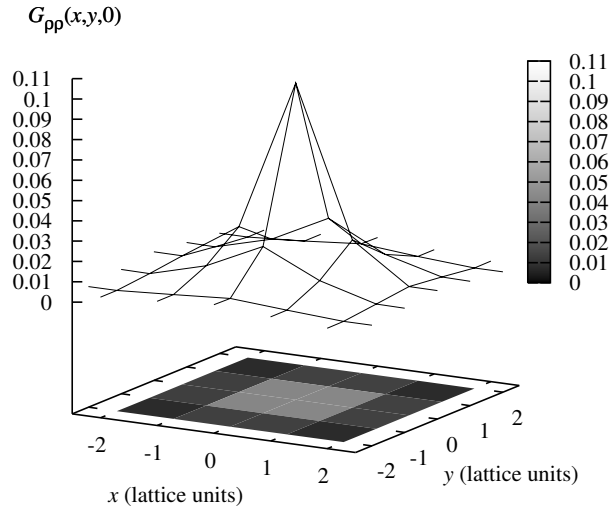}%
\caption{The nucleon density correlation $G_{\rho\rho}(\vec{n}_{s})$ for the
triton in the $xy$-plane at imaginary time $t=0.143$ MeV$^{-1}$.}%
\label{triton}%
\end{center}
\end{figure}

\begin{figure}
[ptbptb]
\begin{center}
\includegraphics[
height=3.026in,
width=3.3875in
]%
{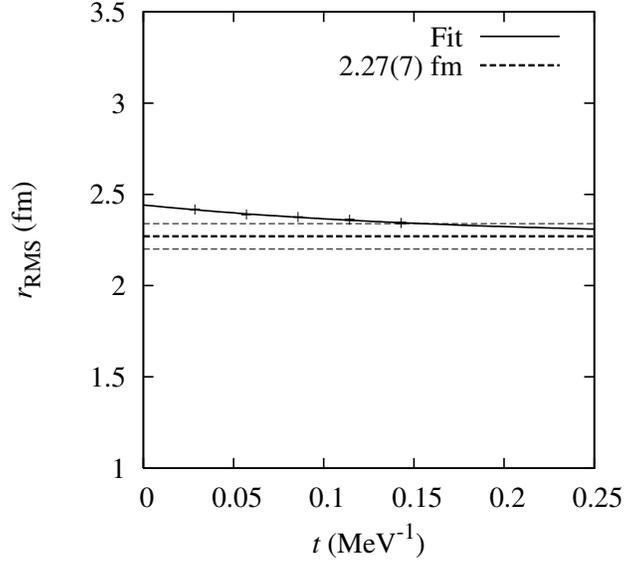}%
\caption{The triton root-mean-square radius for total nucleon density versus
imaginary time $t$.}%
\label{triton_rrms}%
\end{center}
\end{figure}

We also measure the nucleon density correlation $G_{\rho\rho}(\vec{n}_{s})$.
In Fig. \ref{triton} we show $G_{\rho\rho}(\vec{n}_{s})$ in the $xy$-plane at
imaginary time $t=0.143$ MeV$^{-1}$.%
From $G_{\rho\rho}(\vec{n}_{s})$ we can also extract the root-mean-square
radius for the total nucleon density. \ In Fig. \ref{triton_rrms} we show the
triton root-mean-square radius for the nucleon density as a function of
imaginary time $t$. \ We again fit to a decaying exponential form,%
\begin{equation}
r_{\text{RMS}}(t)\rightarrow r_{\text{RMS}}+c^{\prime}e^{-\Delta E^{\prime}t},
\end{equation}
and extract an asymptotic value of $2.27(7)$ fm. \ This is about $30\%$ larger
than the experimental value of $1.755(9)$ fm for the root-mean-square
radius\ of the electric charge \cite{Amroun:1994qj} and the point proton 
root-mean-square radius 1.60 fm \cite{Carlson:qn}.
Similar values for the triton root-mean-square radius 
are also obtained in the pionless framework \cite{Platter:2005sj}.%

\section{Results for helium-4}

For our simulations for helium-4 we use the same lattice parameters$\ a=(100$
MeV$)^{-1}$, $a_{t}=(70$ MeV$)^{-1}$, $C_{^{3}S_{1}}=-4.780\times10^{-5}$
MeV$^{-2}$, $C_{^{1}S_{0}}=-3.414\times10^{-5}$ MeV$^{-2}$, and $b=0.6$. \ We
again set $L=5$, $L_{t_{o}}=8$ and vary $L_{t_{i}}$ from $2$ to $10$. This
time the standing waves comprising $\left\vert \Psi_{Z,N}^{\text{free}%
}\right\rangle $ are%
\begin{align}
\left\langle 0\right\vert a_{i,j}(\vec{n}_{s})\left\vert \psi_{1}%
\right\rangle  &  =L^{-3/2}\delta_{i,0}\delta_{j,1},\nonumber\\
\left\langle 0\right\vert a_{i,j}(\vec{n}_{s})\left\vert \psi_{2}%
\right\rangle  &  =L^{-3/2}\delta_{i,0}\delta_{j,0},\nonumber\\
\left\langle 0\right\vert a_{i,j}(\vec{n}_{s})\left\vert \psi_{3}%
\right\rangle  &  =L^{-3/2}\delta_{i,1}\delta_{j,1},\nonumber\\
\left\langle 0\right\vert a_{i,j}(\vec{n}_{s})\left\vert \psi_{4}%
\right\rangle  &  =L^{-3/2}\delta_{i,1}\delta_{j,0}.
\end{align}
This corresponds with a helium-4 system with $J_{z}=0,J=0$ and total momentum
zero. \ For each value of $L_{t_{i}}$ we again produce about $5\times10^{6}$
hybrid Monte Carlo trajectories using $2048$ processors running independent trajectories.

\begin{figure}
[ptb]
\begin{center}
\includegraphics[
height=3.026in,
width=3.3875in
]%
{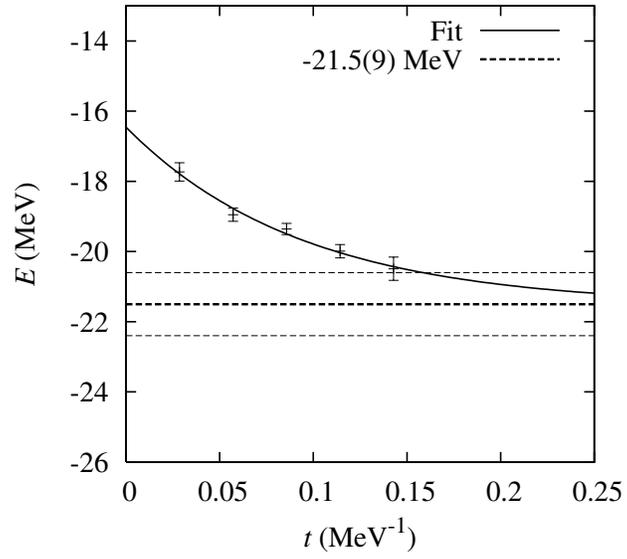}%
\caption{The transient energy for helium-4 as a function of imaginary time
$t$.}%
\label{helium_energy}%
\end{center}
\end{figure}

\begin{figure}
[ptbptb]
\begin{center}
\includegraphics[
height=3.6288in,
width=3.3857in
]%
{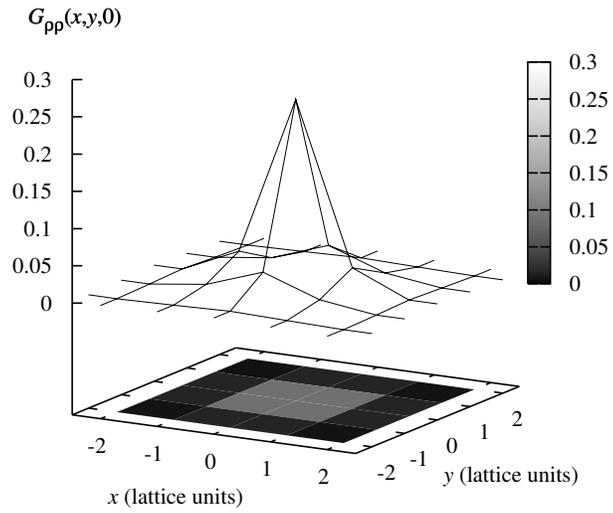}%
\caption{The nucleon density correlation $G_{\rho\rho}(\vec{n}_{s})$ for
helium-4 in the $xy$-plane at imaginary time $t=0.143$ MeV$^{-1}$.}%
\label{helium4}%
\end{center}
\end{figure}

\begin{figure}
[ptbptbptb]
\begin{center}
\includegraphics[
height=3.026in,
width=3.3875in
]%
{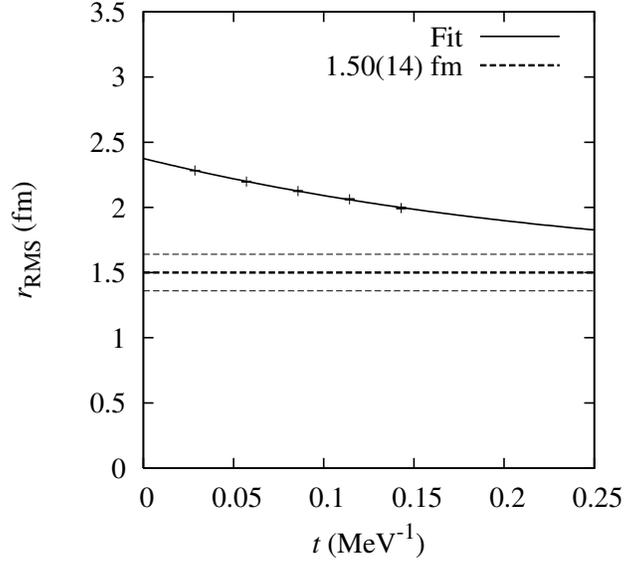}%
\caption{The helium-4 root-mean-square radius for total nucleon density versus
imaginary time $t$.}%
\label{helium_rrms}%
\end{center}
\end{figure}

In Fig. \ref{helium_energy} we show the energy for helium-4 as a function
of imaginary time.%
If we fit $E(t)$ to a decaying exponential form at large $t$ we find an
asymptotic value of $-21.5(9)$ MeV. \ This is about $25\%$ smaller in magnitude than the
experimental result of $-28.296$ MeV. \
Note that we compare directly with the experimental value and have not
corrected for small corrections due to electromagnetic effects.
\ In Fig.~\ref{helium4} we show the
nucleon density correlation $G_{\rho\rho}(\vec{n}_{s})$ in the $xy$-plane at
imaginary time $t=0.143$ MeV$^{-1}$.%
From $G_{\rho\rho}(\vec{n}_{s})$ we compute the root-mean-square radius for
the total nucleon density, and in Fig. \ref{helium_rrms} we show the helium-4
root-mean-square radius for the total nucleon density as a function of
imaginary time $t$. \ Fitting to a decaying exponential form we find an
asymptotic value of $r_{\text{RMS}}=1.50(14)$ fm.%
This is within $10\%$ of the experimentally observed value of $1.673(1)$\ fm
for the root-mean-square radius for the electric charge \cite{Borie:1978} and the point proton 
root-mean-square radius 1.47 fm \cite{Carlson:qn}.

\section{Discussion}

\subsection{Room for more improvement}

The lattice calculations in this study used leading-order chiral effective
field theory with improved contact interactions of the form%
\begin{equation}
f(\vec{q}_{s}\,^{2})\left[  -\frac{1}{2}C\alpha_{t}\rho^{a^{\dag},a}(\vec
{q}_{s})\rho^{a^{\dag},a}(-\vec{q}_{s})-\frac{1}{2}C_{I}\alpha_{t}%
\sum_{I=1,2,3}\rho_{I}^{a^{\dag},a}(\vec{q}_{s})\rho_{I}^{a^{\dag},a}(-\vec
{q}_{s})\right]  ,
\end{equation}
where%
\begin{equation}
f(\vec{q}_{s}\,^{2})=f_{0}^{-1}\exp\left[  -b%
{\displaystyle\sum\limits_{l_{s}=1,2,3}}
\left(  1-\cos q_{l_{s}}\right)  \right]  ,
\end{equation}%
\begin{equation}
f_{0}=\frac{1}{L^{3}}\sum_{\vec{q}_{s}}\exp\left[  -b%
{\displaystyle\sum\limits_{l_{s}=1,2,3}}
\left(  1-\cos q_{l_{s}}\right)  \right]  .
\end{equation}
The three unknown constants $C,C_{I},b$ were determined using the deuteron
binding energy, $^{1}S_{0}$ scattering length, and the average effective range
$\frac{1}{2}\left(  r_{0}^{^{1}S_{0}}+r_{0}^{^{3}S_{1}}\right)  $. \ The
scattering lengths and effective ranges were computed using L\"{u}scher's
formula. \ We used only one value for the lattice spacing, $a=(100$
MeV$)^{-1}$ and $a_{t}=(70$ MeV$)^{-1}$. \ The simulations for triton and
helium-4 were done at only one volume, a cubical lattice with length $L=5$ or
$9.85$ fm. \ In future studies both the dependence on the lattice spacings and
finite volume should be investigated in some detail.

Lattice results for the deuteron root-mean-square radius and quadrupole moment
agree with experimental values at the $5\%$ level. \ The lattice calculation
of the triton binding energy agrees with experiment to within $5\%,$ while the
triton root-mean-square radius is larger by about $30\%$. \ For helium-4 we
find the binding energy is about $25\%$ smaller than experiment, while the
root-mean-square radius agrees within $10\%$. \ This is clear improvement over
the first attempt with zero-range contact interactions which led to a
clustering instability in the four-nucleon system. 
However, in order to make a more rigorous statement
we also have to consider the full set of
interactions which arise at next-to-leading order in chiral effective field
theory. \ We will discuss this procedure in future work.

\subsection{Computational scaling}

The lattice transfer matrix algorithm has several subroutines which scale
differently with the number of nucleons $A$, the spatial volume $L^{3}$, and
the total number of time steps $L_{t}=2L_{t_{o}}+L_{t_{i}}$. \ Multiplication
of the transfer matrices at each time step on the one-particle states scales
as $A\times L^{3}\times L_{t}$. \ Constructing the $A\times A$ matrix
$\mathcal{M}(\pi_{I}^{\prime},s,s_{I},t^{\prime},t)$ from the inner products
of the one-particle states scales as $A^{2}\times L^{3}$, while calculating
the determinant of $\mathcal{M}(\pi_{I}^{\prime},s,s_{I},t^{\prime},t)$ scales
as $A^{3}$ using $LU$ decomposition. 
\ The molecular dynamics trajectories for the hybrid Monte Carlo
updates involves computing the derivatives of $H_{HMC}$ with respect to
$p_{\pi_{I}^{\prime}},p_{s},$ $p_{s_{I}},\pi_{I}^{\prime},s,s_{I}$. \ The
calculation of these derivatives require various loops which scale as
$L^{3}\times L_{t}$, $L^{6}\times L_{t}$, and $A^{2}\times L^{3}\times L_{t}.$

With the same lattice parameters used in our triton and helium-4 simulations,
we investigate the computational scaling for the transfer matrix algorithm as
a function of the number of nucleons $A$. \ We use $L=5$, $L_{t_{o}}=8$,
$L_{t_{i}}=10$ and the free particle standing waves%
\begin{align}
\left\langle 0\right\vert a_{i,j}(\vec{n}_{s})\left\vert \psi_{1}%
\right\rangle  &  =L^{-3/2}\delta_{i,0}\delta_{j,1},\nonumber\\
\left\langle 0\right\vert a_{i,j}(\vec{n}_{s})\left\vert \psi_{2}%
\right\rangle  &  =L^{-3/2}\delta_{i,0}\delta_{j,0},\nonumber\\
\left\langle 0\right\vert a_{i,j}(\vec{n}_{s})\left\vert \psi_{3}%
\right\rangle  &  =L^{-3/2}\delta_{i,1}\delta_{j,1},\nonumber\\
\left\langle 0\right\vert a_{i,j}(\vec{n}_{s})\left\vert \psi_{4}%
\right\rangle  &  =L^{-3/2}\delta_{i,1}\delta_{j,0},\nonumber\\
\left\langle 0\right\vert a_{i,j}(\vec{n}_{s})\left\vert \psi_{5}%
\right\rangle  &  =L^{-3/2}2^{1/2}\cos\left(  \frac{2\pi n_{z}}{L}\right)
\delta_{i,0}\delta_{j,1},\nonumber\\
\left\langle 0\right\vert a_{i,j}(\vec{n}_{s})\left\vert \psi_{6}%
\right\rangle  &  =L^{-3/2}2^{1/2}\cos\left(  \frac{2\pi n_{z}}{L}\right)
\delta_{i,0}\delta_{j,0},\nonumber\\
\left\langle 0\right\vert a_{i,j}(\vec{n}_{s})\left\vert \psi_{7}%
\right\rangle  &  =L^{-3/2}2^{1/2}\cos\left(  \frac{2\pi n_{z}}{L}\right)
\delta_{i,1}\delta_{j,1},\nonumber\\
\left\langle 0\right\vert a_{i,j}(\vec{n}_{s})\left\vert \psi_{8}%
\right\rangle  &  =L^{-3/2}2^{1/2}\cos\left(  \frac{2\pi n_{z}}{L}\right)
\delta_{i,1}\delta_{j,0}.
\end{align}
The initial state $\left\vert \Psi_{Z,N}^{\text{free}}\right\rangle $ is
composed of all free particle standing waves $\left\vert \psi_{i}\right\rangle
$ with $i\leq A$. \ For $A=2$ this is the deuteron system, for $A=3$ the
triton system, for $A=4$ helium-4, for $A=5$ helium-5, for $A=6$ lithium-6,
for $A=7$ lithium-7, and for $A=8$ beryllium-8. \ We note that for $A>4$ the
momentum of $\left\vert \Psi_{Z,N}^{\text{free}}\right\rangle $ is not exactly
zero but rather a wavepacket with a small spread in momentum centered around
zero momentum.

\begin{figure}
[ptb]
\begin{center}
\includegraphics[
height=3.026in,
width=3.3875in
]%
{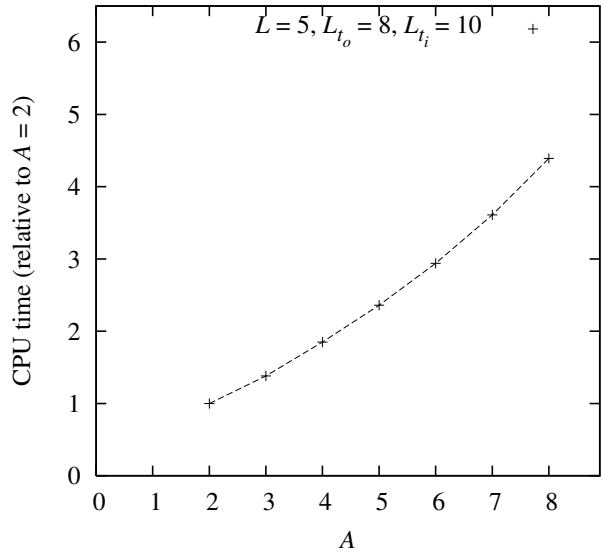}%
\caption{CPU time versus the number of nucleons, $A$, measured relative to the
$A=2$ deuteron system.}%
\label{load}%
\end{center}
\end{figure}

In Fig. \ref{load} we show the CPU times for $A=2,\cdots,8$ relative to the
CPU time for $A=2$ with the same number of hybrid Monte Carlo trajectories.%
We see that for $A\leq8$ the CPU time is approximately linear in $A$. \ This
suggests that the subroutines which scale as $A\times L^{3}\times L_{t}$
dominate the CPU time for the simulation for $A\leq8$.

\begin{figure}
[ptb]
\begin{center}
\includegraphics[
height=3.026in,
width=3.3875in
]%
{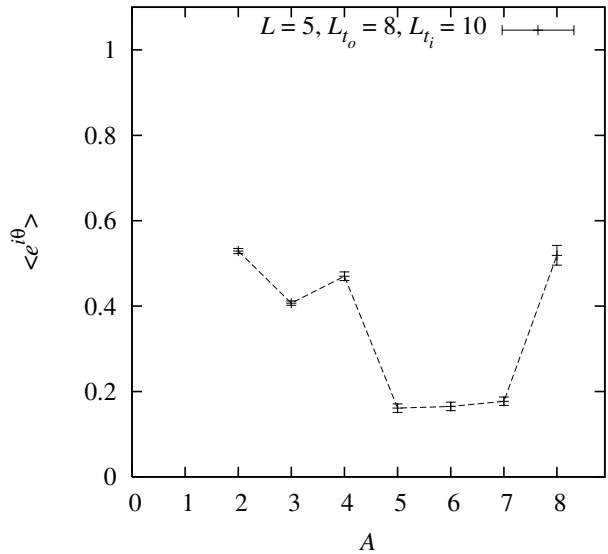}%
\caption{The average phase $\left\langle e^{i\theta}\right\rangle $ versus the
number of nucleons, $A$.}%
\label{phase}%
\end{center}
\end{figure}

In Fig. \ref{phase} we show the average phase $\left\langle e^{i\theta
}\right\rangle $ as defined in Eq.~(\ref{e_i_theta}) versus the number of
nucleons. \ We note the relative maxima in $\left\langle e^{i\theta
}\right\rangle $ at multiples of $4$. \ This can be explained by the
suppression of sign and phase oscillations due to approximate $SU(4)$
symmetry. \ The $SU(4)$ suppression of oscillations is strongest for systems
with equal numbers of spin-up and spin-down protons and neutrons. The results
for the CPU time and average phase indicate that simulations of light nuclei
with $A\leq8$ can in fact be performed without much difficulty using the Monte
Carlo transfer matrix algorithm presented here. \ We plan to pursue these
studies in future work.%

\section{Summary}

We have described simulations of light nuclei on the lattice using a transfer
matrix projection Monte Carlo method at leading order in chiral effective
field theory. \ We included lattice pion fields and auxiliary fields to
reproduce the physics of instantaneous one-pion exchange and the leading-order
S-wave contact interactions. \ To avoid a clustering instability we also
included higher-derivative contact interactions which adjust the S-wave
scattering amplitude at higher momenta. \ There are a total of three unknown
constants, $C,C_{I},b,$ and these were determined using the deuteron binding
energy, $^{1}S_{0}$ scattering length, and the average effective range
$\frac{1}{2}\left(  r_{0}^{^{1}S_{0}}+r_{0}^{^{3}S_{1}}\right)  $. \ We find
agreement between lattice results and experimental data at the $5\%$ level for
all calculated properties of the deuteron. \ The lattice result for the triton
binding energy agrees with experiment to within $5\%$ and the triton
root-mean-square radius is within $30\%$. \ For helium-4 the binding energy is
within $25\%$ while the root-mean-square radius agrees within $10\%$. \ We
expect that the description will improve when higher-derivative operators are
treated systematically by matching phase shifts on the lattice at higher
momentum. \ We have determined that the simulations for light nuclei with up
to eight nucleons can be done without much difficulty using the lattice
methods described here.

\section*{Acknowledgements}

We thank J\"urg Gasser, Hans-Werner Hammer, Christoph Hanhart, Andreas Nogga, Gautam Rupak,
Thomas Sch\"{a}fer, Ryoichi Seki, Bira van Kolck, and Wolfram Weise
for useful discussions. \ In particular we thank
Andreas Nogga for help with supercomputer access and discussions which led to
the functional form used here for the improved contact interactions. \ Partial
financial support of Deutsche Forschungsgemeinschaft (grant BO 1481/6-1), by
BMBF (grant 06BN411),  U.S. Department of
Energy (grant DE-FG02-03ER41260) 
and by the Helmholtz Association (contract number VH-NG-222)
is gratefully acknowledged. \ This research
is part of the EU Integrated Infrastructure Initiative Hadronphysics under
contract number RII3-CT-2004-506078. \ The computational resources for this
project were provided by the Forschungszentrum J\"{u}lich. \ D.~L. thanks the
Forschungszentrum for partial financial support and the University of Bonn for
their kind hospitality.

\bibliographystyle{apsrev}
\bibliography{LO4_02}

\end{document}